\documentclass{aa}
\usepackage[stable]{footmisc}
\usepackage{amsmath}
\usepackage{mathcomp}
\usepackage{txfonts}
\usepackage{placeins}
\usepackage[colon]{natbib}
\bibpunct{(}{)}{;}{a}{}{,}  % to follow the A&A style
\usepackage{graphicx}
\usepackage{epstopdf}
\usepackage{ifthen}
\usepackage[breaklinks, colorlinks, citecolor=blue]{hyperref}
\usepackage{overpic}
\usepackage{fixltx2e}
\usepackage{color}

\usepackage{rotating}

\usepackage[table,usenames,dvipsnames]{xcolor}
\usepackage[absolute]{textpos}
\usepackage[latin1]{inputenc}

\usepackage{etoolbox}
\makeatletter
\patchcmd\@combinedblfloats{\box\@outputbox}{\unvbox\@outputbox}{}{%
   \errmessage{\noexpand\@combinedblfloats could not be patched}%
}%
 \makeatother

\newcommand{\lsst}{{LSST}} 
\newcommand{\bdt}{BDT} 
\newcommand{\bdth}{{\sc bdt~80\%}} 
\newcommand{\bdtn}{{\sc bdt~90\%}} 
\newcommand{\gold}{{\it LSST golden selection cut}}
\newcommand{\cube}{{\it BigCube}}
\newcommand{\zp}{photo-z}
\newcommand{\zs}{spectro-z}
\newcommand{\col}{\color{black}}

\begin{document}
%\title{Impact of photometric redshifts on the LSS power spectrum and the BAO scale determinations in the LSST survey}
%\titlerunning{Impact of photometric redshifts on the LSS power spectrum in the LSST survey}
\title{Impact of photometric redshifts on the galaxy power spectrum and BAO scale in the LSST survey}
\titlerunning{Impact of photometric redshifts on the galaxy power spectrum and BAO scale in the LSST survey}
\subtitle{}
\author{
Reza Ansari\inst{1}
\and Adeline Choyer\inst{2}
\and Farhang Habibi\inst{1}
\and Christophe Magneville\inst{3}
\and Marc Moniez\inst{1}
\and St\'ephane Plaszczynski\inst{1}
\and C\'ecile Renault\inst{2}\thanks{corresponding author}
\and Jean-St\'ephane Ricol\inst{2}
\and Julien Souchard\inst{2}}
\institute{LAL, Univ. Paris-Sud, CNRS/IN2P3, Universit\'e Paris-Saclay, Orsay, France \and Univ. Grenoble Alpes, CNRS, Grenoble INP, LPSC-IN2P3, 38000 Grenoble, France
\and DSM/Irfu/SPP, CEA-Saclay, F-91191 Gif-sur-Yvette Cedex, France}
\date{Received DD MM 2018/ Accepted DD MM 2019}

\abstract 
{Imaging billions of galaxies every few nights during ten years, LSST should be a major contributor to precision cosmology in the 2020 decade. High precision photometric data will be available in six bands, from near-infrared to near-ultraviolet. The computation of precise, unbiased, photometric redshifts up to $z=2$, at least, is one of the main LSST challenges and its performance will have major impact on all extragalactic LSST sciences.}
{We evaluate the efficiency of our photometric redshift reconstruction on mock galaxy catalogs up to z=2.45 and estimate the impact of realistic photometric redshift (hereafter \zp) reconstruction on the large-scale structures (LSS) power spectrum and the baryonic acoustic oscillation (BAO) scale determination for a LSST-like photometric survey. We study the effectiveness of the BAO scale as a cosmological probe in the LSST survey.}
{We have performed a detailed modelling of the \zp~distribution as a function of galaxy type, redshift and absolute magnitude using our \zp~reconstruction code with a quality selection cut based on a Boosted decision tree (\bdt). We simulate a catalog of galaxies in the redshift range $[0.2-2.45]$ using the Planck 2015 $\Lambda$CDM cosmological parameters over 10,000 square-degrees, in the six $ugrizy$ bands, assuming LSST photometric precision for a ten-year survey. The mock galaxy catalogs are produced with several redshift error models. The LSS power spectrum is then computed in several redshift ranges and for each error model. Finally we extract the BAO scale and its uncertainty {\col using only the linear part of the LSS spectrum}.}
{We have computed the fractional error on the recovered power spectrum $\sigma[P(k)]/P(k)$ which is dominated by the shot-noise at high redshift ($z \gtrsim 1$), for scales $k \gtrsim 0.1$,  due to the \zp~damping. The BAO scale can be recovered {{\col with a percent or better accuracy level}} from $z = 0.5$ to $z = 1.5$ using realistic \zp~reconstruction. Outliers ({\it i.e.} galaxies with catastrophic \zp) can represent a significant fraction of galaxies at high redshift ($z \gtrsim 2$), causing bias and errors on LSS power spectrum measurement.}
{Reaching the LSST requirements for \zp~reconstruction is crucial to exploit the LSST potential in cosmology, in particular to measure the LSS power spectrum and its evolution with redshift. Although the BAO scale is not the most powerful cosmological probe in LSST, it can be used to check the consistency of the LSS measurement. {\col Moreover we show that the impact of \zp~smearing on the recovered isotropic BAO scale in LSST should stay limited up to $z \approx 1.5$, so as long as the galaxy number density balances the \zp~smoothing}.}


%%%%%%%%%%%%%%%%%%%%%%%%%%%%%%%%%%%%%%%%%%%%%%%%%%%%%%%%%%%%%%%%%%%%%%\bibliography{biblio}
%%%%%%%%%%%%%%%%
%%%%%%%%%%%%%%%%%%%%%%%%%%%%%%%%%%%%%%%%%%%%%%%%%%%%%%%%%%%%%%%%%%%%%%%%%%%%%%%%%%%%%%

\maketitle

\section{Introduction}

The six-band ($ugrizy$) Large Synoptic Survey Telescope (\lsst) survey, described in \citet{science_book}, will yield a sample of about ten billion galaxies over a huge volume.
It will be the largest photometric galaxy sample of the next decade to study the large-scale
structures (LSS) of the Universe. It aims to characterise the distribution and evolution of matter
on extragalactic scales through observations of baryonic matter at a broad range of wavelengths.
The LSS encode crucial information about the content of the Universe,
the origin of the fluctuations and the cosmic expansion background in which the structures
evolve. Of particular interest is the imprint on galaxy clustering of baryon acoustic
oscillations (BAO), which reflects the acoustic waves at recombination,
related to the sound horizon at that epoch.
The BAO scale is sufficiently small that it is possible to measure it precisely
with a large volume survey; yet, this scale is large enough that it is not significantly altered by non-linear evolution.
The BAO features can be used as a standard ruler to measure distances
and constrain the dark energy equation of state, especially when used in combination with the weak-lensing.

\lsst ~will observe
every few nights half of the whole sky up to a magnitude of 24. After 10 years of survey, the depth of the stacked
images should reach magnitude 27 in the i-band and LSST should provide
a catalog of a few billions of galaxies with a photometry good enough 
to be useable for cosmology. 
Indeed, doing cosmology implies to accurately know the galaxy
positions. While the angular coordinates will be exquisitely
{\col measured} thanks to the LSST optics
and camera design, it will be much more challenging to reconstruct the
third coordinate, the redshift. 

{\col Results} using spectroscopic data of the BOSS survey \citet{Dawson},
part of the Sloan Digital Sky Survey
III, allowed to measure precisely the BAO scale  and
lead to accurate cosmological parameters, for instance \citet{Anderson}, \citet{Aubourg} or more recently \citet{Salazar}, \citet{BOSS}.
Although BAO constrains with LSST might be not fully competitive with the future
spectroscopic survey DESI \citet{DESI} or later Euclid \citet{Euclid},
it is crucial to perform BAO analysis to cross-correlate with the other LSST
probes. {\col The reader can refer to~\citet{Zhan2018} and references therein for a recent review of the cosmology reach of LSST, 
including the BAO probe}. The sensitivity of each probe to systematic errors and to dark
energy parameters is different, so it is mandatory to check the
consistency of all results and powerful to combine them \citet{Alonso}.
The study of BAO in LSST can in return be a way to test our \zp~reconstruction
and tune the quality cuts to deal with the total number of
galaxies versus the number of outliers. The
current DES survey \citet{Drinkwater}, with 5 filters only, has already
to face similar challenges~\citet{DES}{\col , obtaining a 4~\% distance measurement of the standard ruler~\citet{DES_BAO}}.

The goal of this paper is to test the impact of the use of \zp~to reconstruct the LSS power spectrum, considering 
 several options on the required
quality up to z=2.2. {\col Compared to some previous work on BAO with photometric redshifts in LSST~\citet{Zhan09},~\citet{Abrahamse} where simpler LSST observation and \zp~error model were used, 
we have done a complete analysis using realistic mock galaxy catalogs with galaxies}. The cosmological impact is illustrated
by comparing the BAO scales measured using spectroscopic redshifts (hereafter \zs) or  \zp~of galaxies from a mock catalog; our BAO analysis is
similar to the method described in \citet{Blake}. The BAO
scale extraction is simply based on the Wiggle method described in \citet{Beutler}.

The paper is organised as follows. We first present the LSST survey and the goal of the paper. Section~\ref{sec:simucat} is dedicated to an overview of the mock catalog of galaxies and Sect.~\ref{sec:photoz} presents the photometric redshift reconstruction characteristics and performances. Section~\ref{sec:ps} presents method and results regarding the matter power spectra and the BAO scale. We have developed a Montecarlo simulation to   easily link precision on the recovered BAO scale and \zp~performances presented in Sect.~\ref{sec:discuss}. Finally, we illustrate in Section 6 .We end by summarising our results in
Sect.~\ref{sec:ccl}. 

%%%%%%%%%%%%%%%%%%%%%%%%%%%%%%%%%%%%%%%%%%%%%%%%%%%%%%%%%%%%%%%%%%%%%%%%%%%%%%%%%%%%%%
%%%%%%%%%%%%%%%%%%%%%%%%%%%%%%%%%%%%%%%%%%%%%%%%%%%%%%%%%%%%%%%%%%%%%%%%%%%%%%%%%%%%%%
\section{LSST mock galaxy catalog simulation and analysis procedure overview}
\label{sec:simucat}

\begin{figure}[htpb]
\centering
\includegraphics[width=\columnwidth]{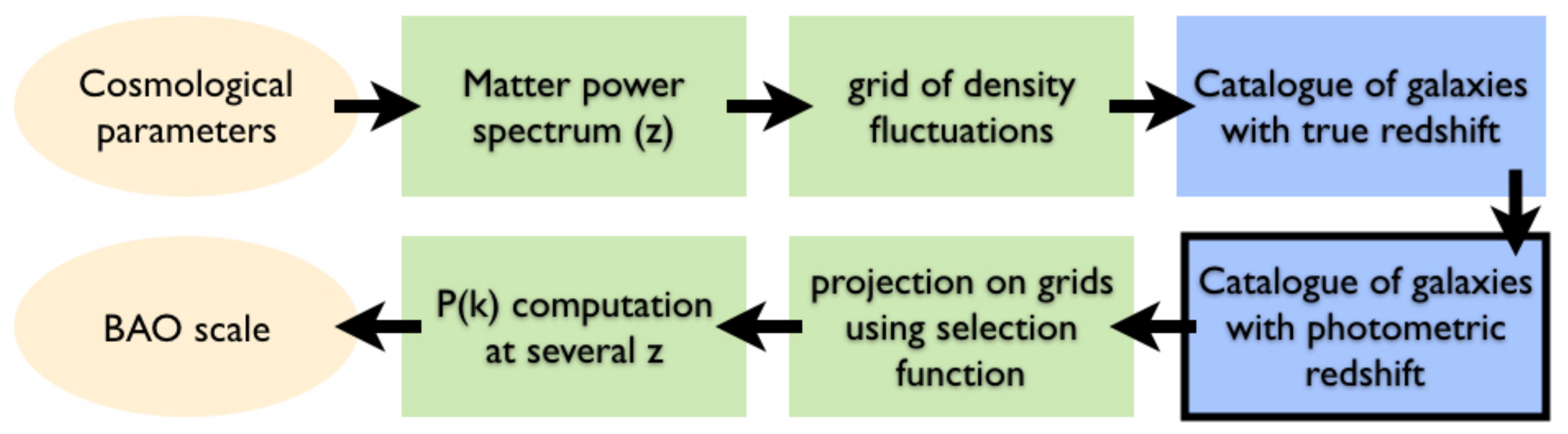}
\caption{Graphic representation of the main steps of the
  simulation chain, from cosmological parameters up to the BAO scale measurement.
} 
\label{fig:scheme_simu}
\end{figure}

A complete end to end simulation and analysis pipeline has been developed  to measure 
the performance of a LSST-like photometric survey for recovering the LSS power spectrum.
Starting from a set of cosmological parameters and going through mock galaxy catalogs with realistic 
\zp, the pipeline is used to extract the BAO scale at several redshifts.
The main blocks of this pipeline are depicted in Fig.~\ref{fig:scheme_simu}. 

We start from the matter power spectrum $P(k)$ computed at a given reference redshift $z_0$, from which a huge grid of density fluctuations $\delta \rho/\rho$ is calculated, through random generation of Fourier coefficients and Fourier transform (FFT). 

The grid of density field is then converted  into a catalog of galaxies using galaxy luminosity functions (LFs). We use a set of 
LFs corresponding to different galaxy classes mapped into spectral types.   
It should be stressed that, in this procedure, we assume that the local galaxy absolute luminosity is directly proportional to the 
local matter density $(1 + \delta \rho / \rho)$.  

The key ingredient for the study presented here is the error model for assigning "observed" redshifts to the galaxies. We
 compare several cases, including a realistic photo-z reconstruction through the 
 FastPz method discussed below.

The mock galaxy catalog is then analysed: after computation of the redshift dependent selection function,
the galaxy angular positions and redshifts are converted into 3D positions using a fiducial cosmology. 
Galaxies are projected into 3D-grids positioned at several redshifts. 
The "observed" power spectrum $P_z(k)$ for each redshift range is computed 
through a Fourier transform, averaging over all Fourier modes within a $k-$bin.
Finally, the BAO scale and its uncertainty are extracted  from the ratio of $P_z(k)$  to a wiggle-less power spectrum through a dedicated fit procedure described in Sect.~\ref{sec:kbao} and compared to the expected value from the input cosmological parameters.  

{\col This pipeline directly provides a catalog, so we ignore all systematics introduced by the catalog extraction from the images, even if effects like blending~\citet{Jones} or dithering~\citet{Awan} affect in different manner photometric and spectroscopic surveys. The simplistic way galaxy catalogs are produced do not allow to derive the precision of the effective reconstruction of the matter power spectrum by LSST. This methodology and the following analyses are however well suited to evaluate the impact of realistic \zp~redshifts, compared to \zs~redshifts, on the matter power spectrum, and more specifically on the reconstruction of the BAO scale.  }

%%%%%%%%%%%%%%%%%%%%%%%%%%%%%%%%%%%%%%%%%%%%%%%%%%%%%%%%%%%%%%%%%%%%%%%%%%%%%%%%%%%%%%
\subsection{Computation of the density field grid}

The aim is to simulate the density field on a large portion of the sky in order to get a simplified view of half of the sky that will be monitored by LSST. We define a volume, hereafter named \cube, which has a comoving depth of 5,600~Mpc along the $Oz$ axis and a transverse comoving section of 
10,000 $\times$ 10,000~Mpc$^2$. 
The grid center is set at 3,200 Mpc from the observer. This \cube~is divided into cells of 
$8 \times 8 \times 8 \, \mathrm{Mpc^3}$, this cell size being small enough compared to the BAO scale ($\sim 150$~Mpc).
Some non-linearities may affect the galaxy distribution around or below the cell scale but, for our purpose, we do not need to take into account such additional complexities. 

The matter power spectrum is initially computed for the redshift of the \cube~center with the cosmological parameters of the fiducial model. We use the set of parameters defined in \citet{planck_cosmo} in the conservative case, based on the TT power spectrum with lensing reconstruction, polarisation only at low multipoles and external data (column 3, Tab. 4): 
the Hubble constant is then $H_0 = 67.90 \, \mathrm{km.s^{-1} .Mpc^{-1}}$, the cold dark matter reduced density is $\Omega_\mathrm{CDM} = 0.2582$, the baryonic matter reduced density is $\Omega_B = 0.0483$ and the (linear) power spectrum is described by a power-law with an amplitude $\sigma_8 = 0.8154$ at a scale of 
$ 8 \, h^{-1} \mathrm{Mpc}$ and a spectral index $n_s = 0.9681$. By hypothesis in the $\Lambda$CDM model, the neutrino mass is {\col minimal} and the geometry is flat. 

Density fluctuations are generated from the power spectrum $P(k)$ using the standard formalism, described for example in 
 \citet{eisenstein}. 
Fourier components $F(\vec{k})$ for the matter density fluctuation field are randomly generated following the  power 
spectrum $\langle | F(\vec{k}) |^2 \rangle = P(k=|\vec{k}|)$. The matter density fluctuation field $\delta \rho / \rho $ 
is then computed through an FFT on the array of generated Fourier components. 
We next apply the growth factor to each cell in the cube, according to its redshift. This method is less accurate than
the one implemented in CLASS~\citet{class} but the resulting spectra differs only marginally. 
{\col This simplified method is sufficient, as our aim is mainly to quantify the loss of information induced by the \zp~reconstruction on the matter power spectrum and as we do not try in this work to quantify uncertainties on cosmological parameters.}

The density field grid is then clipped, setting cells with $\delta \rho/\rho <-1$ to $-1$ to avoid 
cells with negative matter density. This clipping has no effect on the results presented in this paper, 
as we will use the power spectrum of the clipped grid as the reference power spectrum. We have also checked that the 
distortions of the power spectrum due to clipping, ranging from a factor 0.5-0.6 to 0.8 depending on the k-scale, 
do not significantly affect its overall shape and BAO features.

%%%%%%%%%%%%%%%%%%%%%%%%%%%%%%%%%%%%%%%%%%%%%%%%%%%%%%%%%%%%%%%%%%%%%%%%%%%%%%%%%%%%%%
\subsection{Simulation of the catalog}
\label{selfunc}

The density fluctuations are afterwards converted into galaxy number fluctuations,
assuming proportionality of light to matter density $(1+\delta \rho / \rho)$. The key ingredient for this 
step is the use of observed galaxy LF to populate each grid cell with a number of galaxies 
with absolute magnitudes and galaxy classes or spectral types following the LFs.
Details of the procedure are provided in \citet{gorecki}, the main difference being that we have used 
LF parameters derived by \citet{Zucca} for the study presented here, instead of the one 
derived by \citet{Dahlen} used in our previous work. Both sets of LFs provide the distribution 
of galaxies divided in broad types, in several redshift ranges. We have used the 
Schechter parameters from \citet{Zucca} with fixed power law index $(\alpha)$.
LF parameters are given for galaxies divided in three broad types, labeled as Early, Late or StarBurst,
corresponding roughly to Elliptical, Spiral and Irregular morphological types. These three broad 
categories are then mapped into a library of spectral types.

The main steps for converting the density fluctuation grid to a list of galaxies are summarised below:

\begin{itemize}
\item The total number of galaxies $N$ within our survey volume is computed by integrating the LFs with 
a faint absolute magnitude limit well above the LSST detection threshold at all redshifts. This number $N$ is then used to compute the average 
number of galaxies per \cube~cell.
\item For each cell, the average number of galaxies in the cell $\bar{n}(\vec{x})$ is computed according to the local matter density 
$1 + \delta \rho / \rho$.  The number of galaxies in the cell is then drawn with a Poisson distribution 
with mean $\bar{n}(\vec{x})$. Galaxie positions are uniformly generated within each cell, then their angular positions and 
true redshifts are computed from their positions in the grid.  An absolute magnitude and a type are assigned 
to each galaxy according to the multi-type, redshift dependent LFs. We have chosen to use the \citet{Zucca} parametrisation which, once extrapolated to high redshifts, provides a reasonable evolution of the number of galaxies. 
Only galaxies within a cone with half-opening angle of 60~degrees (defining a solid angle of $\pi$, so covering $\approx$10,000 $ \mathrm{deg^2}$) are kept. The catalog in our simulation properly covers the redshifts from 0.2 to 2.45.
\item We use a library of 51 interpolated SEDs from six main SEDs, composed of 4 SEDs of \citet{CWW}: Elliptical, Sbc, Scd or Im, and the 2 SEDs of \citet{Kinney}: SB3 and SB2. Galaxies from each LF class are then distributed over a subset of these 
51~interpolated SEDs. We then compute the apparent magnitudes in each of the six LSST bands ($ugrizy$)  for each galaxy
present in the simulation volume, taking into account the nominal LSST 10-year photometric errors for each band. {\col The photometric errors will depend on the distribution of visits per filter and on the ability to manage the atmospheric properties during observations, as shown in~\citet{graham}.} Note that these errors have been estimated for point-like sources and are applied to extended sources.
Only galaxies with $m_i < 25.3$, satisfying the so-called \gold,~are  kept for the subsequent 
analysis, allowing a reliable estimate of the \zp.
\end{itemize}

As it will be shown and discussed in this paper, the effective density of galaxies observable in LSST as a function of 
redshift is a crucial ingredient to compute its expected performance for measuring the LSS power spectrum 
and the BAO scale. Many values of the Schechter parameters describing the LFs 
can be found in the litterature \citet{Ramos}, \citet{Dahlen}, \citet{Zucca} for instance.
As expected, LF parameters derived from observations have some uncertainties and one 
finds significant differences when galaxy densities in the LSST survey are computed from different 
set of LFs.

\begin{figure*}[h!]
\centering
\begin{tabular}{c c}
\includegraphics[width=\columnwidth]{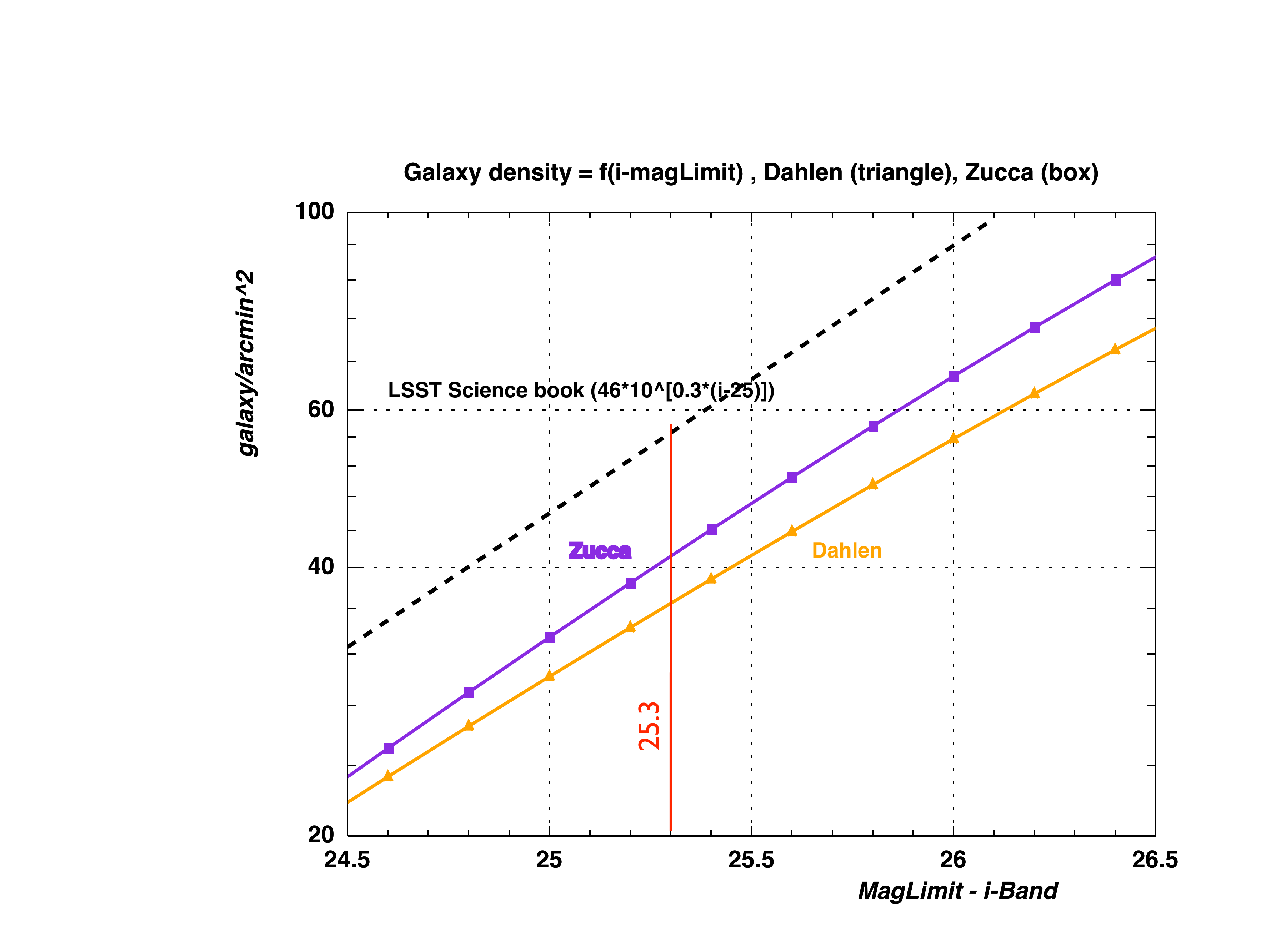} &
\includegraphics[width=\columnwidth]{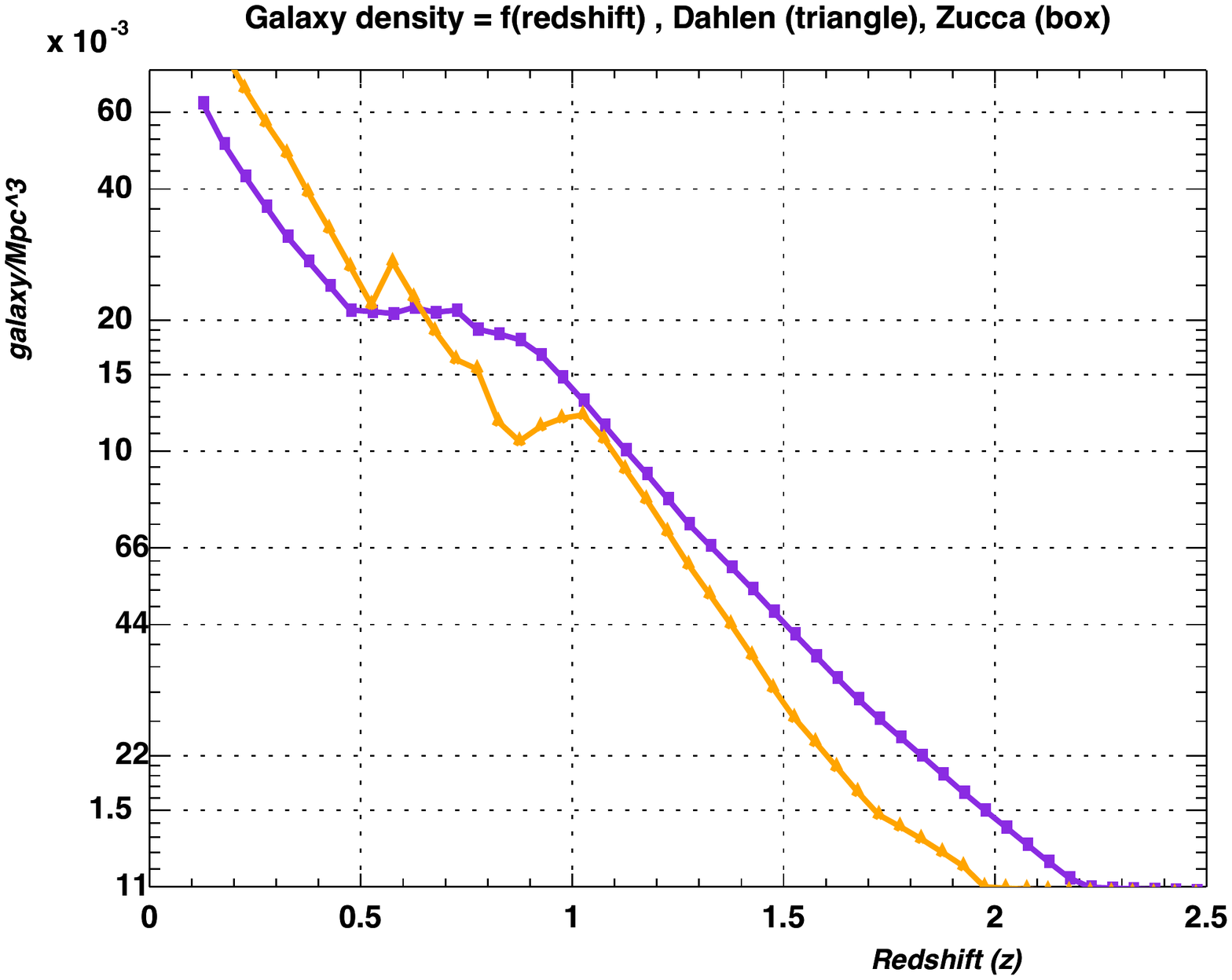} \\
\end{tabular}

\caption{Left: expected galaxy surface density as a function of i-band magnitude limit:
comparison of \citet{science_book} extrapolation, numbers derived from the \citet{Dahlen} and \citet{Zucca}
LFs parameters. Right: number of galaxies per Mpc$^3$  as a function of redshift, 
for the \gold~from \citet{Dahlen} and \citet{Zucca} LFs. Purple squares show results obtained with the  \citet{Zucca} numbers, orange triangles  show results obtained with the  \citet{Dahlen} ones.} 
\label{fig_compare_LF}
\end{figure*}

In the left panel of Fig.~\ref{fig_compare_LF}, we show the expected galaxy surface density number as a function of 
the i-band magnitude limit. We compare the numbers obtained from the extrapolation used in the \citet{science_book} 
and numbers derived from \citet{Dahlen} and \citet{Zucca} LF parameters. 
These curves are obtained by integrating the LF Schechter functions for each redshift,
with the faint end absolute magnitude limit computed from the i-band magnitude limit, taking into account 
the K-correction and the luminosity distance. The observable galaxy density is then integrated as a function 
of redshift, over the cosmological volume element. 
As it has already been pointed out, by \citet{Chang} for instance, the expected galaxy density used in \citet{science_book} 
is rather optimistic. Indeed, for the \gold, the Science book extrapolation yields 
$\sim 55~\mathrm{gal/arcmin^2}$, compared to $\sim 36~\mathrm{gal/arcmin^2}$ for \citet{Dahlen}
and $\sim 41~\mathrm{gal/arcmin^2}$ for \citet{Zucca}. The initial LSST value of $55~\mathrm{gal/arcmin^2}$ seems to come from a too rough first guess.To explain the discrepancy between the two other sets of LFs, we compare their expected number of galaxies per Mpc$^3$ as a function of redshift for the \gold~in the right panel of Fig.~\ref{fig_compare_LF}. The  \citet{Zucca}~LFs lead to a higher galaxy number density at all redshifts above $z\approx 0.6$. So, once integrated up to $z \geq 0.8$, the {\col surface} galaxy number density is higher with this set of Schechter parameters. Moreover the \citet{Zucca}~LFs provides smoother, so more realistic, galaxy density evolution along the cosmic time.

Our simulated galaxy number density should also be compared to the available observed values. The recent Hyper Suprime-Cam Survey has properties rather similar to the expected ones of the LSST survey \citet{HSC}. Their galaxy number density usable for \zp~reconstruction is of almost 22~galaxies per arc-min$^2$, the one usable for weak-lensing is of almost 25~galaxies per arc-min$^2$ and it exceeds  30~galaxies per arc-min$^2$ in some parts of the GAMA15H field \citet{HSC2}. So our galaxy number density sounds reasonable, in agreement with previous LSST forecast for weak-lensing too \citet{Chang}. 
 
From the 28~billions of simulated galaxies, about 1.4 billion satisfy the \gold. This number is roughly a third of the "official" volume of the LSST catalog, firstly because only 10,000 square-degrees have been simulated instead of the 20,000 square-degrees that will be scanned by LSST, secondly because the galaxy density initially foreseen was probably over-estimated,
as we have pointed it out previously.  

\begin{figure}[htpb]
\centering
\includegraphics[width=\columnwidth]{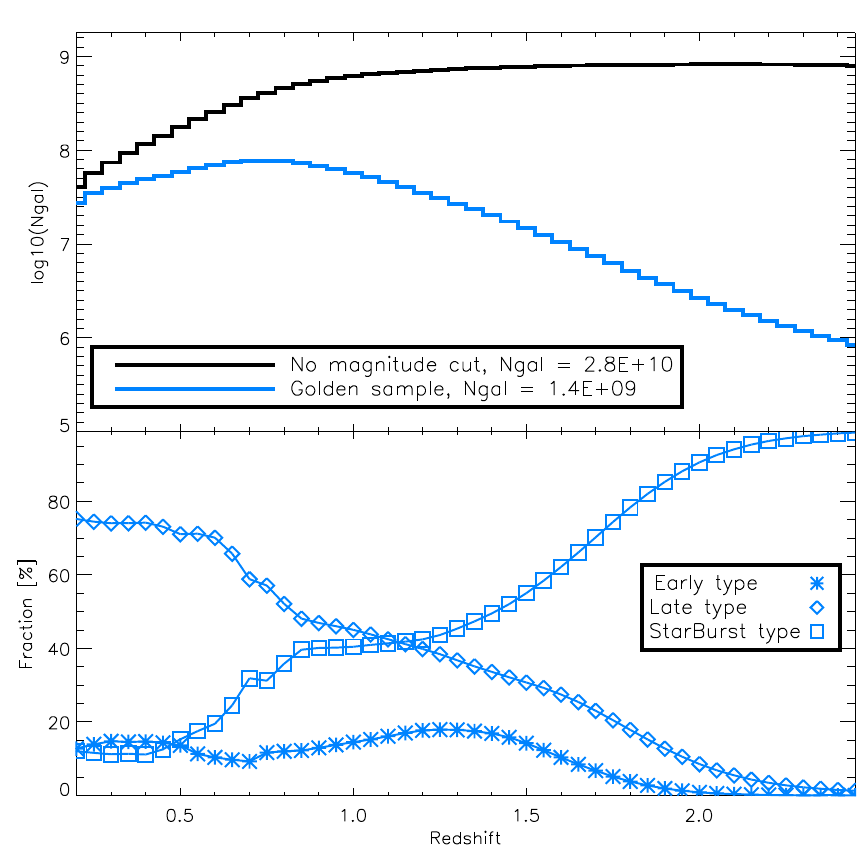}
\caption{Top: Number of galaxies per redshift interval. The \gold ~is defined by $m_i < 25.3$. 
Bottom: relative distribution of the broad types of the galaxies satisfying the \gold~as a function of the
redshift. 
} 
\label{fig:cat_stat}
\end{figure}

Figure~\ref{fig:cat_stat} shows the histograms of the differential number of galaxies as a function of the redshift, before
and after the \gold, as well as the distribution per broad type. Our mock catalog extends from a redshift of 0.2 to a
  redshift of 2.45. About a tenth of the galaxies have a good enough photometry for cosmology purpose at $z
  \approx 1$ and one thousandth  at $z  \approx 2$. 
  
  Note that in the full simulation of the mock catalog, dust extinction, 
photometric errors and use of 51~interpolated SEDs (instead of 6~main SEDs) are included. This comes to be small differences between 
the numbers from the curve labeled {\em Zucca}, derived from the full simulation and the results labeled {\em Zucca} of the simplified model presented in Fig.~\ref{fig_compare_LF}.

In order to take into account realistic uncertainties on the redshift,
random errors are added to the true redshift $z_{\rm s}$. Four error models are
studied:
\begin{itemize}
\item a Gaussian error with $\sigma=0.03*(1+z_{\rm s})$,
\item a more realistic \zp~reconstruction using PDF distributions described in \ref{subsec:photozMethod},
\item a photometric error improved by quality cuts on the \bdt~variable
as described in \ref{sec:bdt}, keeping 90 (\bdtn) or 80~\% (\bdth) of the galaxies.
\end{itemize}

%%%%%%%%%%%%%%%%%%%%%%%%%%%%%%%%%%%%%%%%%%%%%%%%%%%%%%%%%%%%%%%%%%%%%%%%%%%%%%%%%%%%%%
%%%%%%%%%%%%%%%%%%%%%%%%%%%%%%%%%%%%%%%%%%%%%%%%%%%%%%%%%%%%%%%%%%%%%%%%%%%%%%%%%%%%%%
\section{Photometric redshift reconstruction}
\label{sec:photoz}

%%%%%%%%%%%%%%%%%%%%%%%%%%%%%%%%%%%%%%%%%%%%%%%%%%%%%%%%%%%%%%%%%%%%%%%%%%%%%%%%%%%%%%
To guarantee the cosmological capabilities of \lsst, 
requirements on the quality of the \zp~reconstruction
are provided in the \citet{science_book}.
Statistical 
properties of the distribution of $e_z = (z_s  - z_p)/(1+z_p)$ as a function of $z_p$ where $z_s$ 
denotes  the \zs~and $z_p$ denotes the \zp, have to verify the following requirements:
\begin{itemize}
\item the interquartile range IQR of $e_z$, is lower than 0.05 (the goal is lower than 0.02),
\item the fraction of outliers $f_{\rm out}$ defined as the fraction of galaxies with $e_z  > 0.15 $ is lower than 10~\%,
\item the bias $b$ defined as the median value  of $e_z$ is lower than 0.003.
\end{itemize}

Note that our Gaussian error model corresponds to IQR~$\approx 0.04$ as IQR $\approx 2 \times 0.6745 \times \sigma$.

\subsection{Method}
\label{subsec:photozMethod}

The \zp~reconstruction is performed with a template fitting method presented in details
in \citet{gorecki}.
The SED library is made of 51 templates used for the photometry
computation. Note that this step of the work is expected to be rather optimistic
for LSST redshift reconstruction performance since our SED library completely represents  the simulated observations.
The method uses the prior proposed in \citet{Benitez} to remove some degeneracies (mostly Lyman / Balmer breaks)
 based on magnitude distribution in the i-band.
For each galaxy we have to compute a 3D-grid of
$\chi^2(z,T,E_{\rm B-V})$ where the redshift $z$,
the galaxy type $T$ and the extinction from inner dust $E_{\rm B-V}$ are
the three parameters determined by the \zp~reconstruction code. The extinction follows the 
 \citet{Cardelli} or \citet{Calzetti} law, depending on the
galaxy type.
We marginalise the $\chi^2$ grid over the two other parameters to get the marginalised 1D
posterior probability density functions $P(z)$, $P(T)$ and $P(E_{\rm B-V})$.
The redshift estimator can be chosen as $z_p^\chi$
(value of the triplet that minimise the $\chi^2$ in the grid)
or $z_p^{marg}$, value that maximises the posterior probability density function $P(z)$.
In this work we use the latter one: $z_p = z_p^{marg}$.
This code gives \zp~performances similar to other
public codes (Ricol et al., in preparation) but allows us more freedom, in particular the possibility to compute our PDF functions.

%%%%%%%%%%%%%%%%%%%%%%%%%%%%%%%%%%%%%%%%%%%%%%%%%%%%%%%%%%%%%%%%%%%%%%%%%%%%%%%%%%%%%%
\subsubsection{Outliers rejection}
\label{sec:bdt}

The originality of the method is the outliers effective rejection with a
machine learning technique based on the characteristics of the PDFs.

We have improved this technique in the present work. The Likelihood Ratio method presented in \citet{gorecki} has been indeed
replaced by a Boosted Decision Tree (\bdt) algorithm based on the same discriminant parameters but more robust
and performant.
The parameters used in the \bdt~are:
\begin{itemize}
\item the number of peaks in the marginalised 1D posterior probability density functions
  denoted by $N_{pk}(\theta)$, where $\theta$ is either $z$, $T$ or
  $E_{\rm B-V}$,
\item when $N_{pk}>1$, the logarithm of the ratio between the height of the secondary peak over
the primary peak in the 1D posterior probability
  density functions, denoted $R_L(\theta)$,
\item when $N_{pk}>1$, the ratio of the probability associated with the secondary
  peak over the probability associated to the primary peak in the 1D posterior probability
  density functions, denoted by $R_{pk}(\theta)$
(the probability is taken proportional to the integral of the PDF between two minima either side),
\item the absolute difference between $z_{pk}$ and $z_{pk}^{marg}$,
  denoted by $D_{pk} = |z_{pk} - z_{pk}^{marg}|$,
\item the maximum value of $\log(\mathcal{L})$ where $\mathcal{L}$ is the likelihood,
\item the colors $\vec{C} = (u-g,g-r,r-i,i-z,z-y)$,
\item the \zp~value $z_p = z_{pk}^{marg}$.
\end{itemize}

The \bdt~method is trained on an independent sample of 100,000 galaxies. The impact of the training sample (number of galaxies, completeness)
has been studied in details in \citet{gorecki} and shows no significant
bias within reasonable statistics.

After training, the \bdt~method provides
a \bdt~value between -1 and 1 for each test sample galaxy, outliers having lower \bdt~values than good reconstructions.
We can then apply a cut on this number.

The technics on \zp~reconstruction are continuously improved, see for
instance~\citet{Suveges}, \citet{Cavuoti}, \citet{Sadeh}, \citet{Zahra}, \citet{Pasquet}. However, we do not expect significant changes 
in the main features of the reconstructed \zp~distributions and the impact of the LSS 
power spectrum reconstruction.

\subsubsection{Fast \zp~computation}
\label{sec:fpz}
For a sake of computing time we have developed a fast photometric
redshift reconstruction (hereafter FastPZ) tool that allows to quickly
compute the \zp~for each galaxy.
The FastPZ tool is based on true \zp~distributions sorted in bins of true redshift, absolute magnitude and broad type.
We use 121 bins in absolute magnitude $M$ (from -24 to -12 with a step of 0.1), 3 bins in broad type $BT$ (Early, Late or StartBurst)
and 61 bins in redshift $z$ (from 0 to 3 with a step of 0.05) leading to a total of 21,780 bins.
In each bin we simulate several thousands of photometric data,
compute the \zp~for galaxies
satisfying the \gold~and construct the $P(z_p-z_s | M, BT, z_s)$ distributions.
For very low statistic bins we increase the number of galaxies in order to get reliable distributions.

The full method uses photometric information in the six filters. It is the one applicable to real data. In a simulation framework, the fluxes are first computed from the broad type, the absolute magnitude and the true redshift of each galaxy. The FastPZ method directly uses the 3D $(BT, z_s)$ mapping to retrieve the reconstructed \zp~PDF from the broad type and the absolute magnitude of the galaxy.
This method is able to compute 10~millions of \zp's in 1~second while it takes several hundreds of hours with the full method, resulting in a gain of CPU of about 1~million.
 
We have checked that performances of the FastPZ tool and the whole
redshift reconstruction are very similar (Ricol et al., in preparation). 

\begin{figure*}[h!]
\centering
\begin{tabular}{c c}
  \includegraphics[width=\columnwidth]{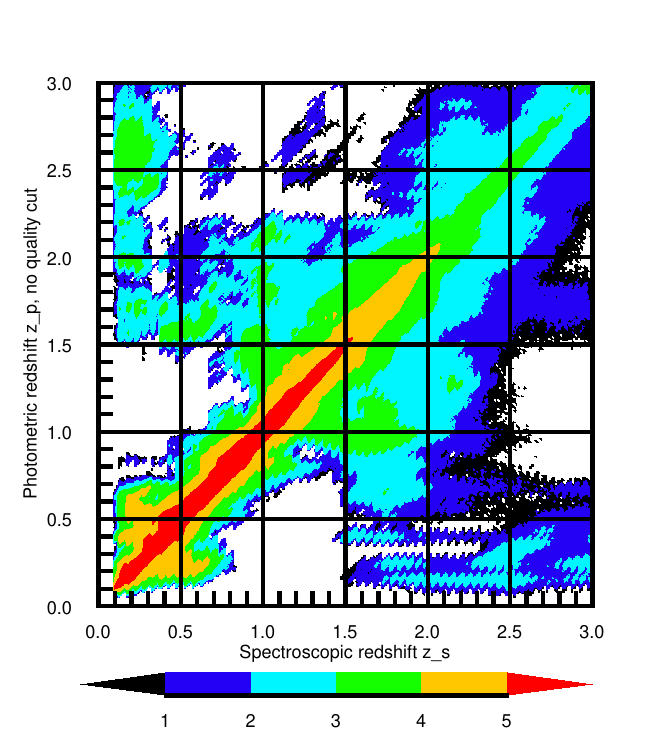}&
  \includegraphics[width=\columnwidth]{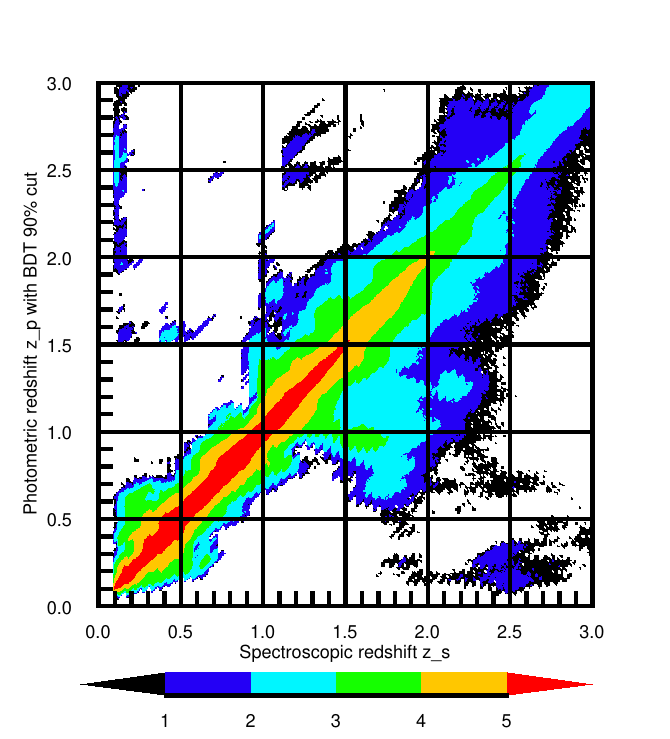}\\
  \includegraphics[width=\columnwidth]{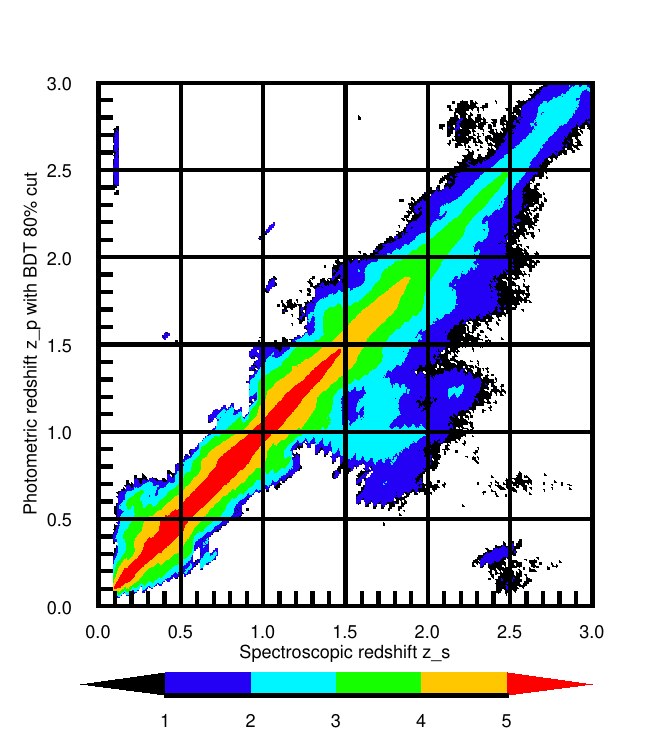}&
\hspace{-1cm}  \includegraphics[width=\columnwidth]{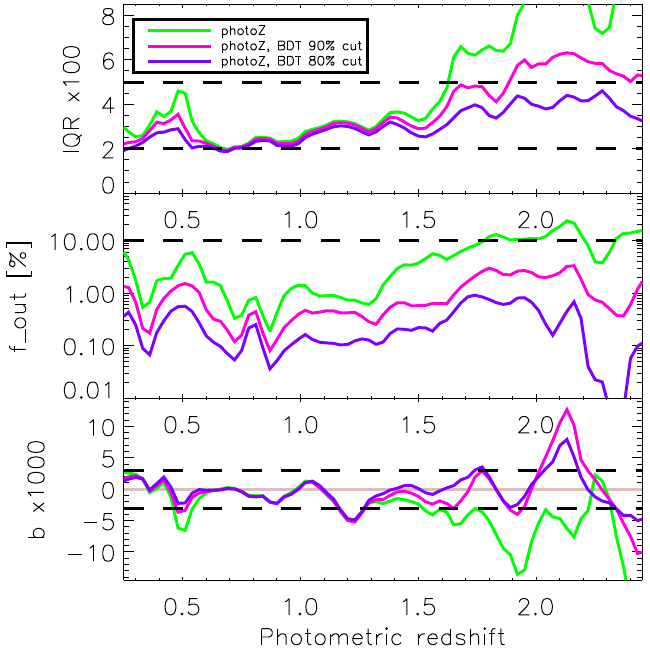}
\end{tabular}
\caption{Distribution of the \zp~$z_{\rm p}$ as a function of the \zs~$z_{\rm s}$; top left:
  \zp~from the FastPZ method (see sect.~\ref{sec:fpz}), top right: with
  the \bdtn~cut, bottom left: with
  the \bdth~cut. The
  colour scale for the galaxy density is logarithmic. 
Bottom right: statistical
  properties of $e_z =(z_s - z_p) / (1+z_p)$  as a
  function of  $z_p$ with (top) the interquartile range IQR of $e_z$,
  (middle) fraction of outliers $f_{\rm out}$, defined by $|e_z| > 0.15$, (bottom)
  bias $b$, defined as the median of $e_z$. Green lines show the values obtained with the
  \zp's without any quality cut while the pink
  (resp. purple) lines show the results with the
  \zp~satisfying \bdtn~(resp. \bdth) cut.
   The dashed black lines correspond to the LSST requirements (plus the goal value in the IQR case). }
\label{fig:zs_zp}
\end{figure*}

Figure~\ref{fig:zs_zp} presents the results of the FastPZ and the performances of the outliers rejection
with the \bdt~method applied on our mock catalog of
$\approx$1.4~billion galaxies.
The \bdt~cut value is tuned to keep a given percentage of the total number of galaxies.

Figure \ref{fig:zs_zp} (top left) shows the density of \zp~$z_{\rm p}$ as a function of the \zs~$z_{\rm s}$.
While most of the reconstructed redshifts lie close to the diagonal,
the plane is almost filled by some ``catastrophic'' reconstructions
({\it i.e.} galaxies labelled as outliers).
Nevertheless, the galaxy density is shown with a logarithmic colour scale, so
 at least one thousand times more galaxies lie in the diagonal area
than in the green areas.

Figure~\ref{fig:zs_zp} (top right and bottom left) shows the same distribution as previously,
but after use of the \bdtn~or \bdth~cut to reject outliers.
Galaxies lying in the diagonal
area are mostly unaffected while all the space above the diagonale is
now almost completely cleared out. The part below the diagonal is much
cleaner too. If the \bdt~cut is always efficient, it
is particularly powerful at high redshift ($z_{\rm s}>1.5$) where the
number of galaxies  with ``catastrophic''
redshift is reduced by at least 2~orders of magnitude.

The quality of the \zp~reconstruction is evaluated
with the estimators used by the LSST
collaboration. These statistical properties of $e_z = (z_s  - z_p)/(1+z_p)$ are plotted as a function of the
\zp~in Fig.~\ref{fig:zs_zp} (bottom right). 
The LSST requirements are more or less satisfied by all the \zp~reconstructions up to $z_{\rm p} \approx 1.6$. For higher redshifts, a \bdt~cut is mandatory, the interquartile range (IQR) remains close the requirement and the bias varies significantly. Note that a similar degradation of the \zp~reconstruction performance is observed above $z=1.6-1.8$ in the real data of the Hyper Suprime-Cam Subaru Strategic Program Data Release 1 \citet{HSC3}.

%%%%%%%%%%%%%%%%%%%%%%%%%%%%%%%%%%%%%%%%%%%%%%%%%%%%%%%%%%%%%%%%%%%%%%%%%%%%%%%%%%%%%%
\subsection{Error model statistics, selection functions and quality cuts}

\begin{figure}[htpb]
\centering
\includegraphics[width=\columnwidth]{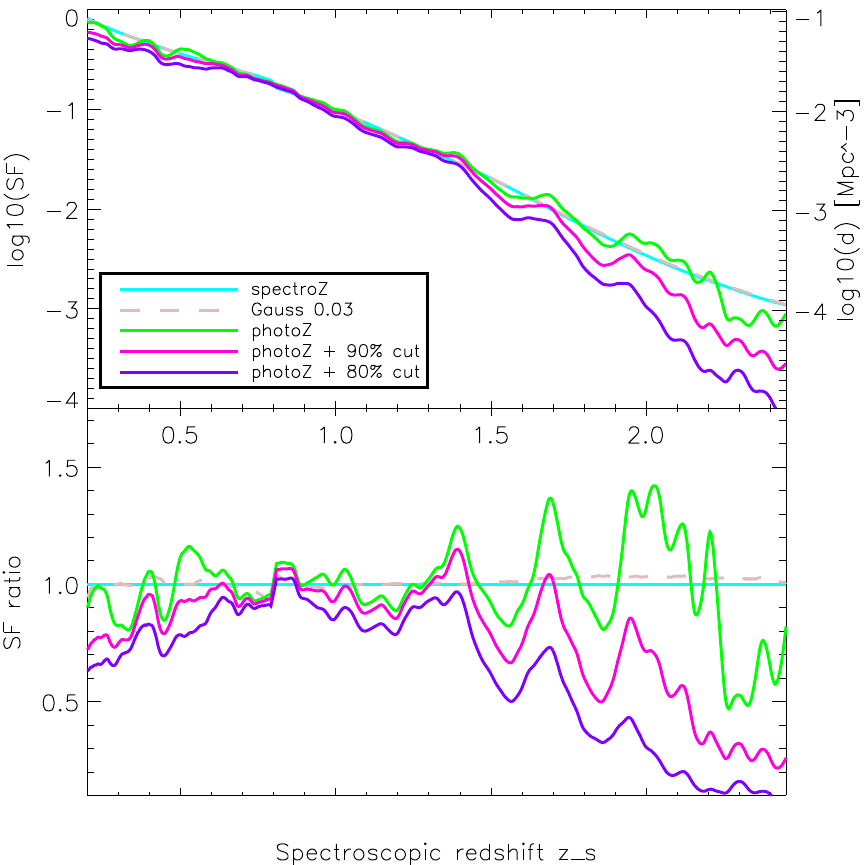}
\caption{Selection functions SF in the redshift range
 [0.2-2.45], using the various estimations of the redshift.
Top: selection functions defined as the ratio between the number of galaxies
  in the catalog (passing the \gold) and the number of simulated galaxies. The
  right axis gives the galaxy
  density $d$ per Mpc$^{-3}$. Bottom:
  ratio between the selection functions and the selection function in
  the spectroscopic case. } 
\label{fig:selfunc}
\end{figure}

Selection functions are computed as the ratio between "observed"
galaxy redshift distributions (spectroscopic, Gaussian or
photometric), satisfying the \gold,~and the "true"
redshift distribution of all galaxies  (black line of Fig.~\ref{fig:cat_stat}). 
 Of course we have to limit the magnitude when "true" galaxies are simulated. "No magnitude cut" means in fact $m_i < 27$, which should be close to the depth of the stacked images after the 10-year survey completion. Galaxies are simulated with absolute magnitude in the range [-24, -13]. This range is appropriate to our simulation as, even at the lower redshifts, the selection function remains below 1 but is close to 1, so we simulate enough, but not too many very faint galaxies.

\begin{figure}[htpb]
\centering
\includegraphics[width=\columnwidth]{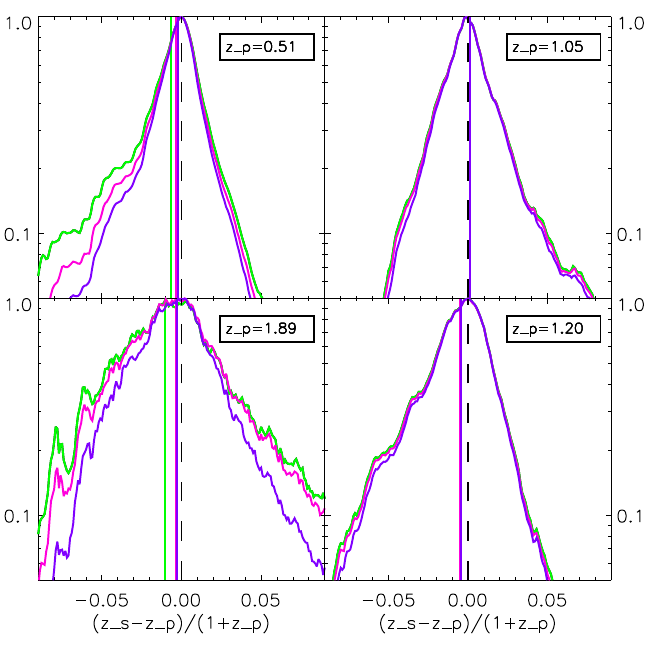}
\caption{Normalised histogram of $e_z =(z_s - z_p) / (1+z_p)$ for 4 bins in \zp. The coloured vertical lines show the value of the bias for the 3 error models on the \zp~considered (see Fig.~\ref{fig:selfunc} for colour legend) while the dashed black line is the reference at 0.} 
\label{fig:bias}
\end{figure}

The selection functions are shown in Fig.~\ref{fig:selfunc}.
The main shape, shown by the cyan line, directly reflects the \gold. 
 The selection function
with the Gaussian error model is very close to the one using \zs, but the ratio of the two functions exhibits an expected slight warp as the smoothing of a
power-law-like distribution leads to a flattening of this distribution.
Cases with \zp~error models are very different: the ratio between the selection function of any of the \zp~cases and the \zs~case exhibits significant structures. Asymmetric $e_z$ distributions are responsible for these visible distorsions since they induce migration of galaxies from true redshift distribution into a distorted measured redshift distribution. It is particularly visible around $z=0.5$ (see Fig.~\ref{fig:bias}, top left). Difficulties in the \zp~reconstruction with asymmetric distributions around this redshift are rather generic, for instance it has been deeply studied by the SDSS collaboration \citet{Beck}.

Additionally, catastrophic redshift reconstructions significantly affect the selection function at high redshift, let's say above $z=1.6$. Indeed, a small fraction (a few percents) of the large low-redshift galaxy population can significantly contaminate (up to 30~\%) the much less populated regions at high redshift. The use of the \bdt~cut makes the selection function slightly smoother as the IQR, the bias and the number of catastrophic \zp are
significantly reduced. The cost is an additional decrease of the number
of usable galaxies, from 2~$\tcperthousand $ to less than 1~$\tcperthousand $ at $z=2$ for instance.

It is interesting to pay attention on the origin of the bias. A residual bias can be corrected for only if it is attributed to a determination method that is known to systematically over or underestimate a measurement. It appears not to be the case here,  as illustrated by Fig~\ref{fig:bias}. The normalised histograms of $e_z$ have been plotted for 4 bins in redshift chosen for their various bias values (visible as the gap between the dashed black and the coloured vertical lines). The $e_z$ distributions have their maximum near zero, but when they are strongly asymmetric, the means differ significantly from the median. Here, applying global shifts to the \zp~~distribution would be improper. Around $z_p$=0.51 or 1.89, the bias in the \zp~case is far beyond the requirements but the \bdt~cut significantly decreases the asymmetry of the $e_z$ distribution and the bias becomes marginally compatible with the requirements at the cost of a significant loss of galaxies. At the opposite, at $z_p$= 1.05 or 1.20, the bias is small, close to the requirement, and the IQR and the fraction of outliers are satisfactory (see Fig.~\ref{fig:zs_zp}): there is clearly a systematic effect but it remains compatible with the requirements and the \bdt cut is not able to improve the symmetry of the $e_z$ distributions.

%%%%%%%%%%%%%%%%%%%%%%%%%%%%%%%%%%%%%%%%%%%%%%%%%%%%%%%%%%%%%%%%%%%%%%%%%%%%%%%%%%%

\subsection{Number density of galaxies}

The redshift evolution of the galaxy number density (in [Mpc$^{-3}$]), for galaxies satisfying the \gold, with and without \bdt~quality cut is shown in top panel of Fig.~\ref{fig:selfunc} (right vertical axis). Additionally, the average galaxy number density for few redshift ranges is given in~Tab.~\ref{tab:stat_ngal_arcmin}.
The impact of the error model on the galaxy density is driven by
the \bdt~cut: 10 or 20~\% of the galaxies are lost with
the \bdtn~or \bdth~cuts, by definition. One can notice the differences on the number of galaxies between the first two redshift bins, using Gaussian or \zp~redshifts: we retrieve the partial redistribution of the galaxies around $z=0.5$, already seen in the shape of the selection functions, when \zp~are used.

\begin{table}
\begin{tabular}{l|c|c|c|c||c}
 $z$ range   & 0.2-0.5 & 0.5-1.0 & 1.0-1.5 & 1.5-2.5 &   0.2-2.5  \\
  \hline
  \hline
$z = z_{\rm G}$          &   6.7 & 19.0 & 9.3 & 2.5 & 37.5 \\
$z = z_{\rm p}$           &  6.2  & 19.7 & 9.3  & 2.4 & 37.6 \\
$z = z_{\rm BDT90}$  & 5.5  & 18.6 & 8.8 &  1.7&  34.6\\
$z = z_{\rm BDT80}$  &  4.9 & 17.4 &  8.0 & 1.1  &  31.4 \\

  \hline
  \hline
\end{tabular}
\caption{Number of galaxies per arc-min$^2$ per redshift range. The \gold~is always applied.
} 
\label{tab:stat_ngal_arcmin}
\end{table}

Table~\ref{tab:stat_z} gives the fraction of galaxies with relative redshift difference $e_z$
lower than some thresholds. 
The LSST requirement on the fractional \zp~error is 5~\% with a goal of 2~\%. Galaxies with a fractional \zp~error of 15~\% are considered as outliers. The three columns of Tab.~\ref{tab:stat_z} correspond to these three thresholds.

\begin{table}
\begin{tabular}{l||r|r|r}
  $f_{{\rm gal}}$ with & $|e_z| < 0.02 $  & $|e_z| < 0.05$ &  $|e_z| < 0.15$   \\
  \hline
  \hline
$z=z_{\rm G}$ & 49 \% &  90 \%  &  100 \% \\
 $z=z_{\rm p}$   & 61 \% & 85 \% & 97 \% \\
$z=z_{\rm BDT9}$  & 65 \%  & 89 \% & 99 \% \\
$z=z_{\rm BDT8}$  & 68 \% & 91 \% & ~100 \%\\
  \hline
  \hline
\end{tabular}
\caption{Fraction of galaxies $f_{{\rm gal}}$ with $|e_z| = |z_s  - z_p|/(1+z_p)$ lower
  than several thresholds for the three photometric error models.
The threshold of 0.02 is the goal, 0.05 is the requirement ; 0.15
defines the outliers.
} 
\label{tab:stat_z}
\end{table}

%%%%%%%%%%%%%%%%%%%%%%%%%%%%%%%%%%%%%%%%%%%%%%%%%%%%%%%%%%%%%%%%%%%%%%%%%%%%%%%%%%%%%%
%%%%%%%%%%%%%%%%%%%%%%%%%%%%%%%%%%%%%%%%%%%%%%%%%%%%%%%%%%%%%%%%%%%%%%%%%%%%%%%%%%%%%%
\section{Computation on the power spectra at different redshifts}
\label{sec:ps}

Our aim is to quantify the impact of photometric redshifts on the LSS power spectrum $P(k)$ 
and the BAO scale $s_A$ determination in LSST. We have thus chosen to use a  Fourier Transform based procedure, 
which is a rather simple, robust and well tested method to determine the LSS power spectrum. Other methods, such as direct computation of the two-point correlation function are often used when analysing observed galaxy catalogs, ~\citet{Szalay} or~\citet{Anderson} for instance. These methods can indeed be better suited to real data, as they can handle more easily inhomogeneous sampling of space, blind spots due to bright stars for instance, which require the use of masks, but more time consuming than FFT. Nevertheless, difficulties related to Fourier-space analysis can be overcome
\citet{BeutlerFourier}.

Assuming a fiducial cosmology, galaxy angular positions and redshifts are converted into cartesian positions which  are 
used to project the galaxies (weighted by the inverse of the selection function) into a 3D-grid $n_\mathrm{obs}(\vec{r})$. There is a difficulty for this grid-FFT method. Cartesian grids are not well adapted to the spherical 
geometry and a clean definition of the redshift range. We have overcome this difficulty by the use of 
several grids, each set paving the spherical shell in a given redshift range, as described below. 

A Fast Fourier Transform (FFT) is so applied to compute the Fourier modes $F(\vec{k})$  of the grid
$n_\mathrm{obs}(\vec{r})$ and the power spectrum of each normalised, mean subtracted galaxy number density grid:
$$ F(\vec{k}) = \mathbf{FFT} \, \left[ \frac{n_\mathrm{obs}(\vec{r})}{\bar{n}_\mathrm{obs}} - 1 \right], \hspace{5mm}  
P_{{\rm obs}}(k) = \langle |F(\vec{k})|^2 \rangle .$$

The BAO scale is then extracted from the power spectrum through a fit of a damped sinusoid to the ratio 
of the observed power spectrum to a wiggle less (no BAO) power spectrum (see Sect.~\ref{sec:kbao}).

\subsection{Multiple grids covering a redshift range}

First, the  volume of universe covered by the simulated galaxy catalog  is organised as several sets of 3D-grids, each set covering a given redshift range. Euclidian ($x,~y,~z)$ axes are defined with respect to the observer, with the $Oz$ axis corresponding to the line of sight through the center of the field. A given direction $\Omega=(\theta, \varphi)$ is defined by two angles, $\theta$ being the angle with respect to $Oz$ and $\varphi$ the angle of the projection of $\Omega$ on the $xOy$ plane 
with $Ox$ axis. 

The grid thicknesses are a compromise between a reasonably small redshift range, sufficient statistics and large enough volume to be sensitive to low k-values ($k_{min} \sim 0.005$). The margin with respect to the initial \cube~limits must be sufficient to avoid artificial loss of galaxies, as the photo-z errors can be large, especially at high redshifts.

\begin{figure}[htpb]
\centering
\includegraphics[width=\columnwidth]{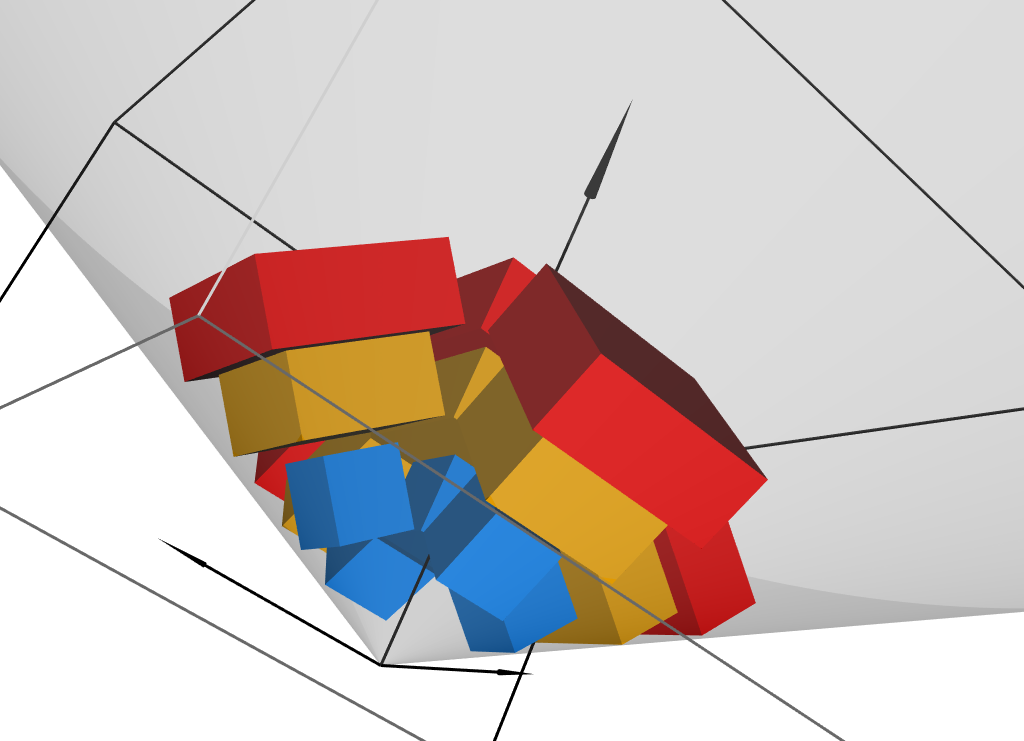}
\caption{Schematic view of the volumes used in the simulation. The black lines mark out the \cube. The grey area shows the
  field of view of $\pi$~sr which is  filled by galaxies  from $z=0.2$ to $z=2.45$. The 5~grids used to compute each power spectrum are drawn in blue, orange or red according to their central redshift.
} 
\label{fig:cube_grid}
\end{figure}

We have defined three sets of 5 grids each, centred at redshifts $z=0.5$, $0.9$, $1.3$, each grid being subdivided  
into $8 \times 8 \times 8$~Mpc$^3$ cells. One grid is centred along the Euclidian $z$ axis
while the others are rotated by 40~degrees in $\theta$ and scattered at 0, 90, 180, 270~degrees in $\varphi$.  
A 3D-view of the grids is drawn in Fig.~\ref{fig:cube_grid}: the initial \cube~is represented by the white box, 
the field-of-view by the grey area while grids are drawn in colours according to their redshift range.

Table~\ref{tab:grids} summarises the main characteristics of these three grid sets. 
After projection of the galaxy catalog into the grids, each cell will contain the number of galaxies 
with a position falling in it, 
weighted by the inverse of the selection function corresponding to the considered error model at the appropriate redshift. 

\begin{table*}
\begin{tabular}{c|c|c|c|c||c|c||c|c||c|c}
 grid   &  cells &  cells &   width,&  volume    &  $N_{\rm gal}$ with  & $<N_{\rm gal}>$  &
  $N_{\rm gal}$  with  & $<N_{\rm gal}>$  &   $N_{\rm gal}$with  &   $<N_{\rm gal}>$  \\
  central  & mean & redshift  & thickness &  [Gpc$^3$]   &   spectroZ  & spectroZ  &      \bdtn  & \bdtn  &   \bdth  &   \bdth   \\
redshift & redshift & range & [$n_{\rm cell}$]  
 & 1  $\rightarrow$ 5 grids  & 
  [10$^6$]   & [cell$^{-1}$]   &
  [10$^6$]   & [cell$^{-1}$]   &
  [10$^6$]    & [cell$^{-1}$]  \\
  \hline
  \hline
 0.5 & 0.51 & 0.36-0.68 & 120, 125 & 0.9 $\rightarrow$ 4.6  & 26.2  & 14.5 & 24.1  & 13.4  & 21.3  & 11.8 \\
 0.9 & 0.93 & 0.72-1.19 & 225, 125 & 3.2 $\rightarrow$ 16.2 & 52.3  & 8.7  & 51.1  & 8.1  & 48.4  & 7.6 \\
 1.3 & 1.36 & 1.08-1.73 & 300, 125 & 5.7 $\rightarrow$ 28.8 & 26.9  &  2.4  & 25.9  & 2.2  & 22.2  & 2.0 \\

 \hline
  \hline
\end{tabular}
\caption{Geometrical description of the grids used for the power
spectrum computation and mean number of galaxies falling in one of the five grids in each set,
with and without \bdt~cut. Grids are composed of $w \times w \times t$ cells of $ 8^3~ \mathrm{Mpc^3}$, with $w$ and $t$ 
the width and thickness values given in the table.  } 
\label{tab:grids}
\end{table*}

\begin{figure}[htpb]
\centering
\includegraphics[width=\columnwidth]{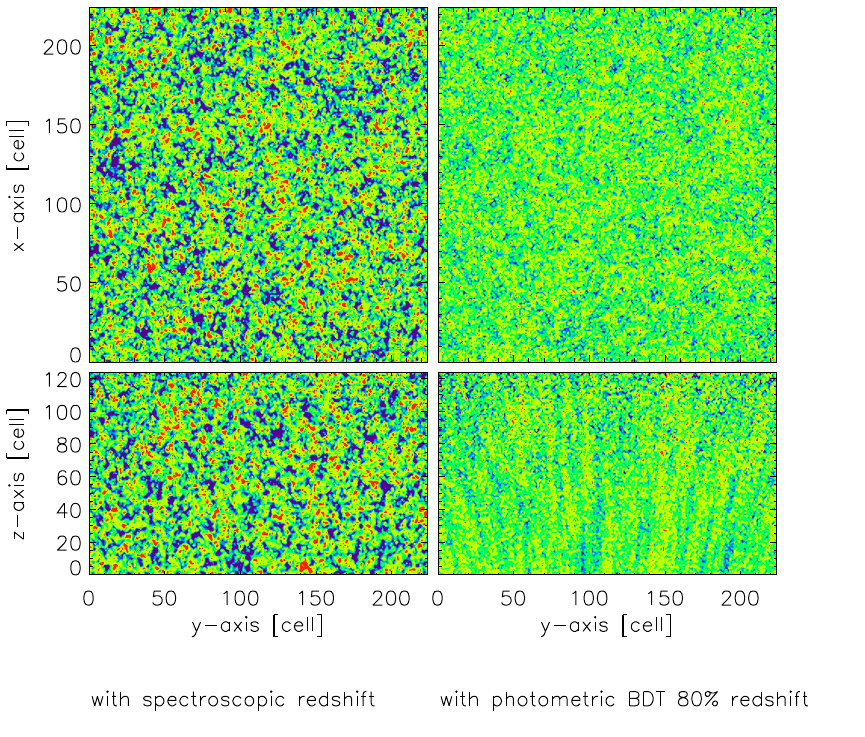}
\caption{Galaxy density contours in slices of the grid centred at
$z=0.9$. Top: the transverse plane $(y,x)$; bottom: the radial plane $(y,z)$. Left : with
  \zs, right : with \bdth~\zp. Slices are  8~comoving Mpc thick (1 cell) and go through the grid center. }
\label{fig:grids_cut}
\end{figure}

Figure~\ref{fig:grids_cut} shows the galaxy density contours of two slices
through the center of the $z=0.9$ central grid. The $(x,y)$ slice 
corresponds to the transverse plane and the $(z,y)$ slice contains the
line-of-sight $Oz$ axis. While density contours are isotropic and small
structures are well contrasted with \zs~(left panels), features appear along the radial (redshift) direction 
and small structures have faded out with \zp~(here \bdth, right panels). 

\subsection{Power spectra and noise estimation}

We have computed power spectra $P_\mathrm{obs}(k)$ for each of the five redshift error models: 
spectroscopic redshifts (no-error), redshifts with gaussian error $\sigma_z = 0.03 (1+z)$ and
photo-z reconstruction without or with 90\% or 80\% BDT cut. 

We also compute the shot-noise contribution by simulation. A separate set of grids is filled by Poisson
noise using the mean galaxy density at the redshift of each cell. Further steps - application
of the error model on the redshift or selection function correction for instance - are then applied as for grids
filled by galaxies. The power spectra of the shot-noise grids $P_\mathrm{SN}$ are flat and the shot-noise
contribution is properly approximated by a constant, as expected, which is determined with a small statistical uncertainty.

The shot-noise subtracted power spectrum is defined by $P_\mathrm{D} (k) = \left\langle P_\mathrm{obs}(k) \right\rangle_{{\rm
3D-grids}} - P_\mathrm{SN} $,
where the subscript $\mathrm{D}$ stands for damped. Indeed, the recovered power spectrum is damped due to redshift 
errors, compared to the underlying galaxy distribution power spectrum.

\begin{figure}[htpb]
\centering
  \includegraphics[width=\columnwidth]{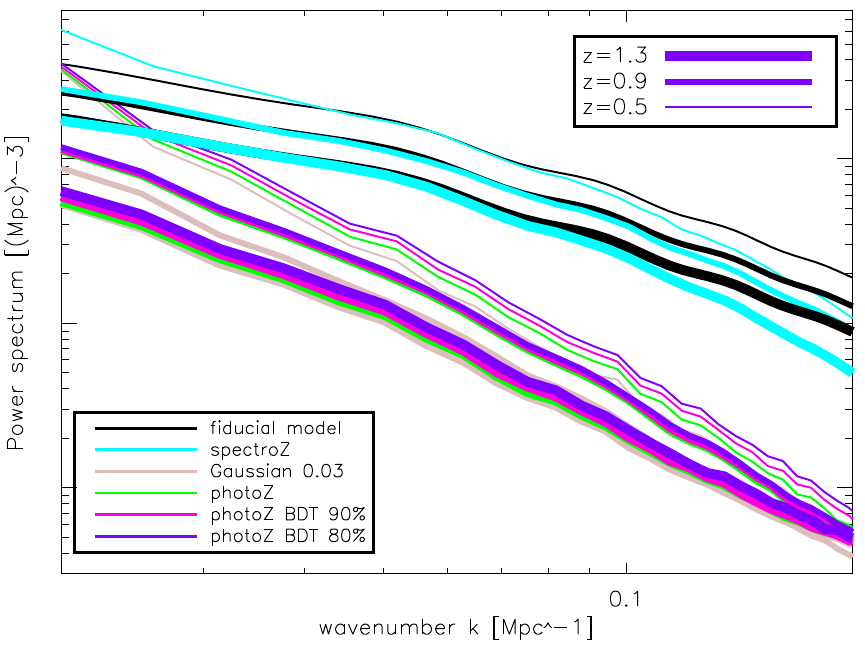}
  \caption{Recovered power spectra $P_\mathrm{D}(k)$ computed from the 5~grids centred at each redshift bin,
   after subtraction of the shot-noise contribution. Black lines correspond to the theoretical (input)
  power spectra while other colours refer to the five redshift error models. The line thickness identifies 
  the grids central redshift, with the  thickness increasing with the redshift. }
\label{fig:ps_ready_fit}
\end{figure}

Theoretical (input) and recovered power spectra from simulated galaxy catalogs, at the three redshifts and for the five redshift error models, 
are shown in figure~\ref{fig:ps_ready_fit}. 

One can see the global decrease in amplitude at all scales when the redshift increases, by comparing for
instance the theoretical shapes (black lines). It is  related to the increasing growth factor with cosmic time. 
The power spectra recovered from catalogs with \zs~for the three redshift ranges  (cyan curves) follow the theoretical
shapes at low $k$, but a moderate damping can be seen, starting around $k=0.1$, which is 
mainly due to the sampling with $8^3~ \mathrm{Mpc^3}$ cells. 

The damping of the power spectrum produced by the photometric redshift smearing is clearly visible 
when errors on redshift are introduced. 
The damping factor reaches a factor $\approx$10 at the BAO scale, around 150~Mpc ($k \approx 0.04$). 

The \bdt~cut reduces the \zp~dispersion, so reduces the damping: the recovered power spectrum amplitude increases with more stringent 
\bdt~requirement. Note that the differences are tiny between the green, pink and purple medium thickness curves,
as \zp~performance is already good around $z=0.9$ without any \bdt~cut. The amplitude of the recovered 
spectra from the grids centred at $z=0.5$ and $z=1.3$ are more sensitive to the \bdt~cut. 

The power spectra for the grids centred at $z=1.3$ start to flatten at high $k$ because they are not far to be shot-noise dominated with an average of only two galaxies per cell (see~Tab.~\ref{tab:grids}).

The statistical errors associated to the recovered power spectra are given by:
\begin{equation} \label{eq:sigma}
\sigma_{P} (k)=
\frac{2 \pi }{k \sqrt{V \delta_k}} \times \left[ P_\mathrm{D}(k) + P_{{\rm SN}} \right]
\end{equation}
where $V$ is the total volume of the grids in a given redshift range and $\delta_k$ is the
sampling width in wavenumber. We have checked that the dispersion of the recovered power spectra from different
mock catalogs do follow the above relation, although with limited number of catalogs, due to CPU and storage intensive 
computations needed to generate and analyse the catalogs.

\begin{figure}[htpb]
\centering
 \includegraphics[width=\columnwidth]{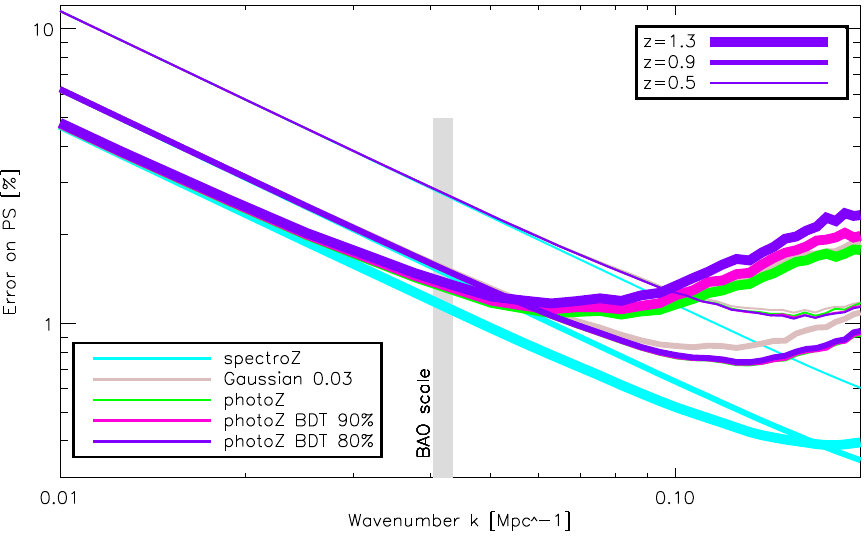}
  \caption{ Fractional statistical error $\sigma_P/P(k)$ 
  of the recovered power spectra $P_\mathrm{D}$ in percent. Colors identify the 
  different redshift error models, while the grid redshifts are distinguished by different line thicknesses. The light gray area shows the wave number corresponding to the BAO scale. 
}
\label{fig:error}
\end{figure}

The fractional statistical uncertainty on the recovered power spectrum $\sigma_P/P(k)$  is plotted in  Fig.~\ref{fig:error}.
It depends naturally on the wavenumber $k$, but also on the redshift interval and on the redshift error model. 
One can distinguish two regimes on these curves:
\begin{itemize}

\item The fractional power spectrum error $\sigma_P/P(k)$ is dominated by the cosmic variance
at low wavenumber ($k<0.015$), for all of the five error models and all redshift intervals. It remains true with spectroscopic redshifts 
up to, at least, $k \approx 0.15$.  The cosmic variance contribution is larger for low redshift as 
the grids are smaller and it evolves as $1/k$.

\item At medium or high wave-numbers,  the fractional statistical error $\sigma_P/P(k)$ is dominated by the shot-noise contribution. The lower limit  in wave number for this regime depends on the redshift range. 

The shot-noise contribution is significantly lower when \zs~are used. Indeed the shot-noise levels, which depend on the mean galaxy density, are very
similar for the spectroscopic, Gaussian and photometric cases. However, their relative value with respect
 to the power spectrum $P_\mathrm{D} (k)$ increase significantly as
 $P_\mathrm{D} (k)$ are damped due to radial smearing (Gaussian or \zp~error models). 

Note that even with \zs, for the 3D-grids centred at $z=1.3$, the relative error contribution flattens at $k \approx 0.2$, as the shot-noise starts to overwhelm the LSS power spectrum. 

\end{itemize}

In summary, we can expect a more accurate BAO scale determination from the grids centred at $z=0.9$, compared to the grids centred at
$z=0.5$ and $z=1.3$. The precision on the recovered power spectrum is limited by the cosmic variance at low redshift ($z=0.5$)
and by the shot-noise at high redshift ($z=1.3$). {\col Note that non-linearities will also soften the oscillations above $k \approx 0.1~\rm{Mpc^{-1}}$}.

%%%%%%%%%%%%%%%%%%%%%%%%%%%%%%%%%%%%%%%%%%%%%%%%%%%%%%%%%%%%%%%%%%%%%%%%%%%%%%%%%%%%%%
%%%%%%%%%%%%%%%%%%%%%%%%%%%%%%%%%%%%%%%%%%%%%%%%%%%%%%%%%%%%%%%%%%%%%%%%%%%%%%%%%%%%%%
\subsection{Extraction of the BAO scale}
\label{sec:kbao}

The baryon acoustic oscillations are subdominant with respect to the global matter
power spectrum shape, as they are hard to see even on theoretical
curves (black lines on Fig.~\ref{fig:ps_ready_fit}). Thereby, the power spectrum, damped by any feature affecting the data or the computation method, follows
a global shape with the small superimposed oscillations. 

We do not want to assume any shape of the damping induced by the smearing produced
by \zp~errors. So we cannot use an analytical
model as it is done for Gaussian error
model \citet{glazebrook}. {\col The appendix contains the description of the procedure that we have developed to estimate the smooth, wiggle-less, power spectrum from the observed one. The oscillating component in the spectrum is extracted, through the fitting of a damped sinusoid, similar to the Wiggle only method~\citet{glazebrook} where the amplitude, the damping scale and the oscillation scale are left as free parameters.}
\begin{figure}[h!]
\includegraphics[width=\columnwidth]{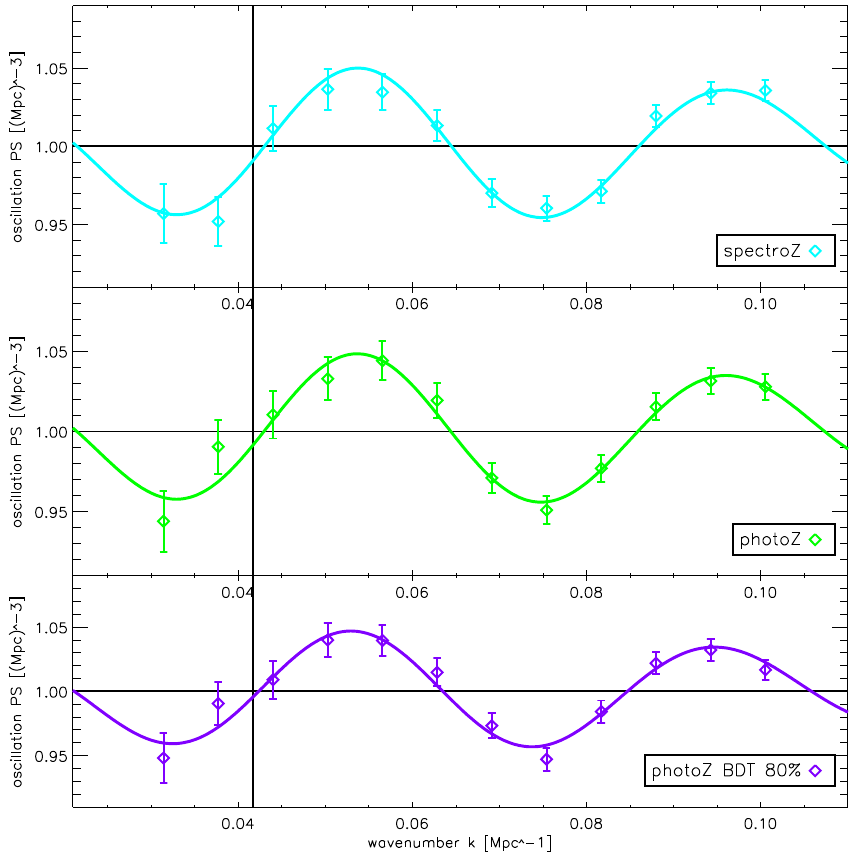}
\caption{{\col Recovered oscillation part of the power spectra in Fourier space for the grids centred at $z=0.9$ considering, from top to bottom, \zs,
  \zp~without and with \bdth~cut.} The coloured lines are the fitted oscillations using the simple
``Wiggle only'' description. The BAO scale is at the first intersection between this fitted damped sinusoid and the horizontal black line. The expected value {\col of $k_A$} from the fiducial Planck 2015 $\Lambda$CDM model (0.0417~Mpc$^{-1}$) is shown by the vertical line.}
\label{fig:osc}
\end{figure}

\begin{figure}[htpb]
\centering
\includegraphics[width=\columnwidth]{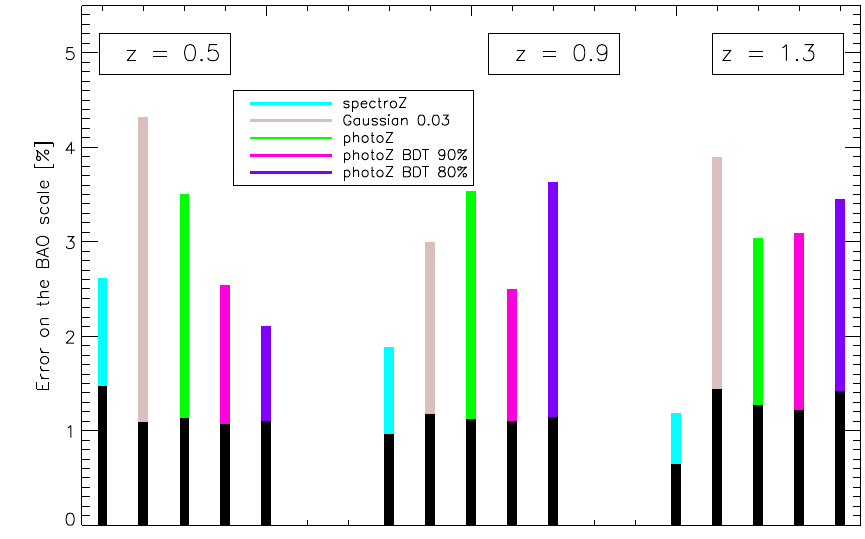}
\caption{Relative errors on the BAO scale fitted from the power spectra for each redshift and each error model. {\col The black bars are the results using the $k$-range [0.03-0.1]~Mpc$^{-1}$ while the color bars  are the results using the $k$-range [0.02-0.07]~Mpc$^{-1}$}.}
\label{fig:psfit}
\end{figure}

{\col As mentioned in Sect.~\ref{sec:simucat}, we have used a simple model in this simulation, ignoring non-linear effects or bias on the LSS power spectrum and mock galaxy catalog generation. Indeed, non-linear clustering affects the power spectrum, leading in particular to a damping of the BAOs features at small scales~\citet{Crocce}, \citet{Rasera}, \citet{Obuljen} or \citet{Seo}. In order to limit over-estimating LSST capability to recover the BAO scale, we have restricted the $k$-range used  to extract  the oscillating component in the power spectrum and to determine the BAO scale $s_a$ . We have used two $k$-ranges, a very conservative one where only $k \leq 0.1~h$/Mpc$^{-1}$ ({\it i.e.} $k \leq 0.07~$Mpc$^{-1}$) have been kept, and a second one, using wave modes up to $k \leq 0.15~h/$Mpc$^{-1}$  ({\it i.e.} $k \leq 0.1/$Mpc$^{-1}$). Indeed  the impact of non linear clustering for low k-modes ($k \leq 0.1~h$/Mpc$^{-1}$) can be safely neglected. However, as one can see for instance from~\citet{Rasera}, the damping of the BAO features due to non-linear clustering is limited up to 0.15~$h$/Mpc$^{-1}$, specially at higher redshifts ($z \geq$ 1), which is more in focus in this work. The comparison of the reconstructed BAO scale error from these two $k$-ranges gives an indication of the amount of information in larger $k$-modes for different redshift bins. 

For illustration purpose, we show the oscillating component  of the LSS power spectrum in the Fourier space, as well as the fitted damped sinusoid, for the redshift $z=0.9$ on Fig.~\ref{fig:osc}.}

The estimated errors on $s_A$ are gathered in Fig.~\ref{fig:psfit} {\col for the  two tested $k$-ranges} . In agreement with previous considerations, results obtained for 3D-grids centred at $z=0.9$ are more precise than results obtained from 3D-grids centred at lower or higher redshifts. {\col The results are two or three times worse if the $k$-modes between 0.07 and 0.1~Mpc$^{-1}$ are removed. Indeed the loss of half of the second oscillation impacts the ability to precisely recover the BAO scale. Nevertheless it does not suppress it as the error remains below 5~\% in all cases}.

The errors are probably underestimated, especially in some cases, so it prevents from proper estimation of the impact of the \bdt~cut on the recovery of the BAO scale. Indeed, the reduced $\chi^2$ are reasonably close to 1 when \zs~are used and for all error models for grids centred at $z=0.9$ and $z=1.3$. But the reduced $\chi^2$ ranges from 5 to 6 when the fit is performed from grids centred at $z=0.5$ with Gaussian or photometric error on the redshift. We interpret this as a hint that another component in the error computation related to the error on the redshift should be included. This component is mainly hidden by the shot-noise at higher redshifts. It is due to the redshift dispersion and not to the bias or the presence of outliers as the $\chi^2$'s are similar with the Gaussian error model (with no bias and virtually no outlier) and with the \zp~error model.

%%%%%%%%%%%%%%%%%%%%%%%%%%%%%%%%%%%%%%%%%%%%%%%%%%%%%%%%%%%%%%%%%%%%%%%%%%%%%%%%%%%%%%
%%%%%%%%%%%%%%%%%%%%%%%%%%%%%%%%%%%%%%%%%%%%%%%%%%%%%%%%%%%%%%%%%%%%%%%%%%%%%%%%%%%%%%
\section{Interpretation and discussion}
\label{sec:discuss}

We will try to explain here some of the general behaviours that we have observed using the full simulation in the previous section. 
We will then use a simple simulation tool implementing these effects to get BAO scale determination uncertainty 
in a LSST-like survey, compared to a fiducial spectroscopic survey. 

The use of \zp~induces two main effects on the galaxy distribution. A first one is the smearing 
along the radial direction, due to reconstructed redshift uncertainties, and a second one is the presence of outliers, 
due to catastrophic redshifts. We can write the observed galaxy number densities $n_{obs}(\vec{r})$ along the line of sight  $\vec{r}$ from which the 
underlying LSS matter power spectrum is computed as:
$$ n_{obs}(\vec{r}) = n_g^{pz}(\vec{r}) + n_{out}(\vec{r})$$ 
where $n_g^{pz}$ (resp. $n_{out}$) denotes galaxy population with reasonable \zp~(resp. outliers).
Introducing the outlier fraction $f_{out}$, we have the following relations between the different average number densities:
$$ \bar{n}_g^{pz} = \left( 1 - f_{out} \right) \, \bar{n}_{obs} ;  \hspace{8mm}  
     \bar{n}_{out} = f_{out}  \, \bar{n}_{obs}. $$
The observed matter distribution power spectrum is computed from the galaxy number density field renormalised
by the average density :
$$ P_{obs}(k) : \mathrm{FFT} \left[ n_{obs}(\vec{r}) / \bar{n}_{obs} - 1 \right]. $$
Neglecting the correlation between galaxies with correctly reconstructed redshifts and outliers, the 
observed power spectrum can be written as:
$$ P_{obs}(k) = \left(1 - f_{out} \right)^2 P_g(k) \times \eta(k) +  f_{out}^2 \, P_{out}(k) + P_{\rm SN}. $$ 

The different terms of this equation correspond to:
\begin{itemize}

\item The  $P_g(k) \times \eta(k)$ term corresponds to the power spectrum of the galaxy number density field radially 
smeared due to \zp~errors, leading to a damping of the power spectrum ($ 0 < \eta(k) \leq 1 $). $ \eta(k) $ follows a scaling law: $\eta(k) = f(k \times \sigma_R )$  with  $\sigma_R$  the standard deviation of the smearing along the radial direction as described in \citet{Blake2003}.

\item Outliers correspond to galaxies from different redshifts shuffled to a large extent, so $  P_{out}(k) \lesssim P_g(k) $; they contribute 
to the overall noise in the observed  galaxy power spectrum. However, the outliers contribution 
would be negligible in most cases, as long as $f_{out} < 10 \%$. 

\item $P_{\rm SN}$ is the combined shot-noise power spectrum from the two populations:
$$  P_{\rm SN} = \frac{ \bar{n}_g }{ \bar{n}^2_{obs} } +  \frac{ \bar{n}_{out} }{ \bar{n}^2_{obs} } = 
    \frac{ \left( 1 - f_{out} \right)  \bar{n}_{obs} }{ \bar{n}^2_{obs} } +  \frac{ f_{out}  \bar{n}_{obs} }{ \bar{n}^2_{obs} } = 
   \frac{1}{ \bar{n}_{obs} } $$
 
\end{itemize}

\begin{figure*}[h!]
\centering
\begin{tabular}{c c}
\includegraphics[width=\columnwidth]{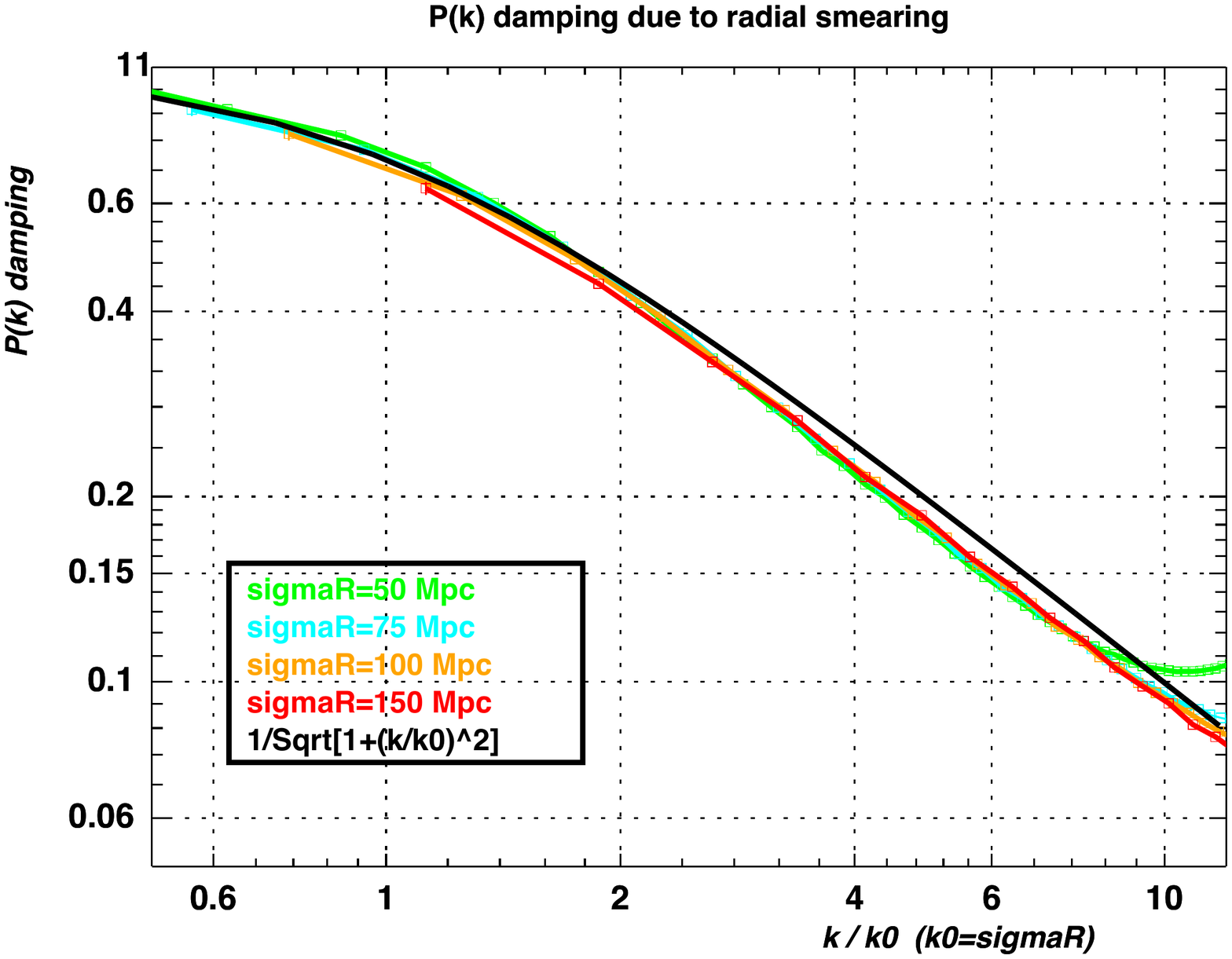} &
\includegraphics[width=\columnwidth]{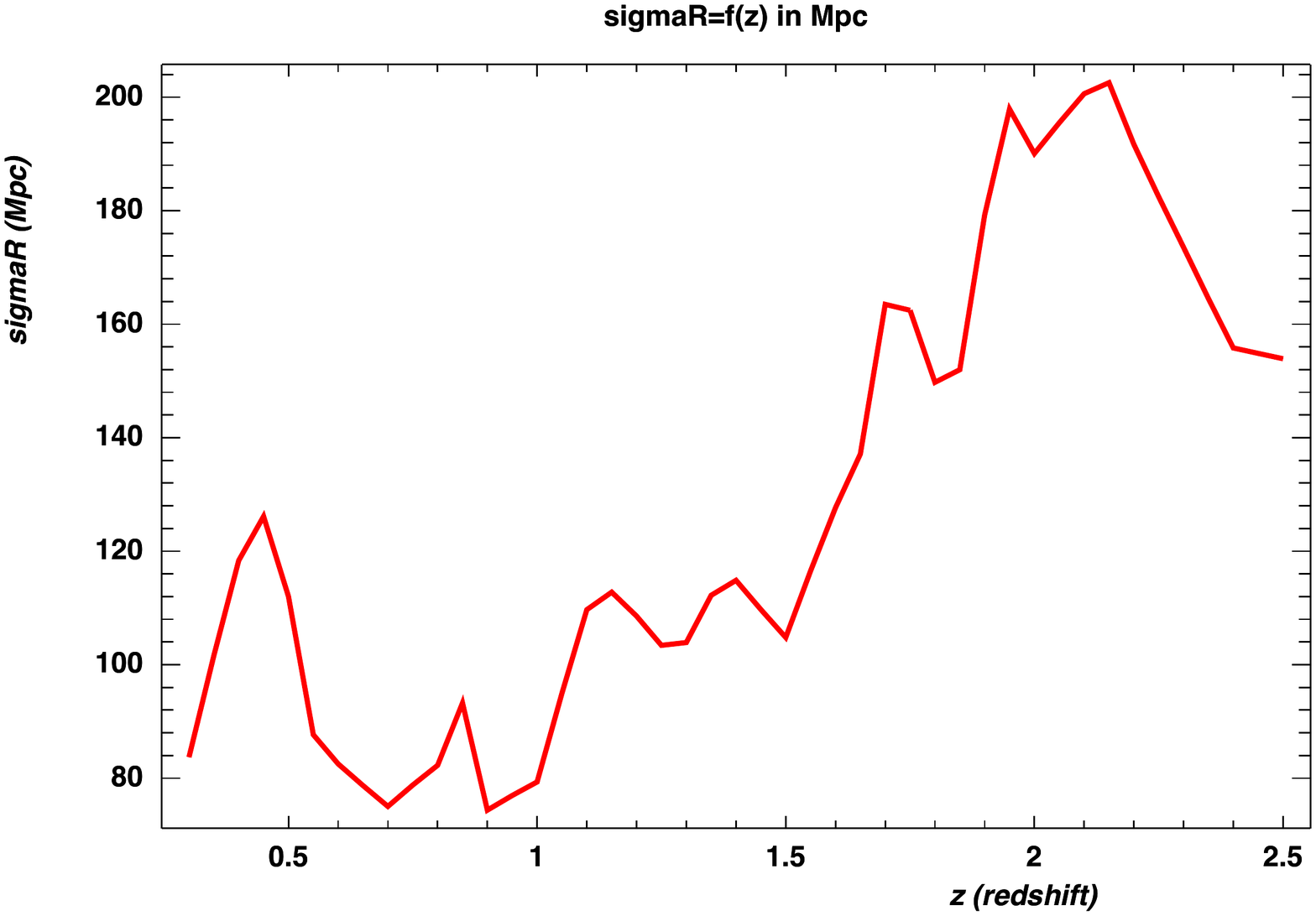} \\
\end{tabular}

\caption{Left: damping function $\eta(k)$ for four values of $\sigma_R = 50, 75, 100, 150 \, \mathrm{Mpc}$ 
(colour curves) and its analytical approximate shape $\left( \sqrt{1 +  \left( k / k_0 \right)^2} \right)^{-1}$ in black.
Right: value of the radial smearing $\sigma_R$  computed from \zp~reconstruction dispersion  
(with \bdtn)  as a function of redshift.  
} 
\label{fig_damping_xsi}
\end{figure*}

We have computed the damping function $\eta(k)$, corresponding to the ratio of the recovered power spectrum 
to the input power spectrum,  for different levels of smearing along the radial direction. 
The left panel of Fig.~\ref{fig_damping_xsi} shows the damping function $\eta(k)$ for four values of 
the standard deviation $\sigma_R = 50, 75, 100, 150 \, \mathrm{Mpc}$ of Gaussian smearing along the radial direction.
The damping function follows very well a scaling law, $\eta(k) = f(k/k_0)$ where $k_0 = \frac{1}{\sigma_R}$,
as shown in Fig.~\ref{fig_damping_xsi}.  The damping function is well represented by the following analytic function
$$ \eta(k) = \frac{1}{\sqrt{1 + \left( k / k_0 \right)^2 }} \hspace{3mm} \mathrm{with} \hspace{2mm} k_0 = \frac{1}{\sigma_R} $$
We have also plotted $\sigma_R(z) = \frac{c}{H(z)} \times \sigma_{pz}(z) $ as a function of redshift 
on the right panel of Fig.~\ref{fig_damping_xsi}, using the redshift dependent $\sigma_{pz}(z)$ 
from full \zp~reconstruction of Sect.~\ref{sec:photoz} with \bdtn~selection cut. 

We have carried a study of the BAO scale reconstruction uncertainty using a toy-program %{\tt (kBAOErr)} 
where 
the observed galaxy power spectrum and associated errors are computed using the expressions given above. 
The galaxy power spectrum has been damped using a redshift dependent radial smearing $\sigma_R(z)$, as represented by the IQR variable in Fig.~\ref{fig:zs_zp}, and a fixed value of 5\% for the outlier fraction. Note that  $\sigma_R(z)$ is very irregular, as the IQR is in Fig.~\ref{fig:zs_zp}; this variability is not due to a lack of statistics but to galaxies spectra features seen by the \zp~method. We have used redshift dependent 
galaxy number density following the values obtained in our full simulation, slightly below the ones 
shown in Fig.~\ref{fig_compare_LF}. We computed  the ratio of the observed power spectrum to a 
theoretical power spectrum without BAO oscillations $R(k) = \left( P_{obs}(k) - P_{\rm SN} \right) / \left( P_{no-osc}(k) * \eta(k) \right)$.
The BAO scale $\mathrm{s_A}$ is then determined by  fitting a damped sinus function 
$A \exp \left( k / k_d \right) \sin \left( \mathrm{s_A} \, k \right) $ to the ratio $R(k)$, with three free parameters: $A, k_d, s_A$,
over {\col the restricted k-range $0.03 \leq k \leq 0.1~\rm{Mpc}^{-1}$, equivalent to $0.04 \leq k \leq 0.15~h\rm{Mpc}^{-1}$}.  {\col A constant quadratic error of 0.4\% has been included to compensate partially for the simplicity of this fitting method.} The increase of the  uncertainty due to \zp~smearing is however marginally sensitive to the fitting procedure. The error is determined from the distribution of reconstructed BAO scales $s_A$.

The survey volumes given in Tab.~\ref{tab:grids} correspond to a survey of 
slightly less than 4,000~deg$^2$. 
{\col Using our toy-program and for $\Omega_{surv}=4,000~\mathrm{deg^2}$, we obtain 
a BAO scale uncertainty $\sigma_\mathrm{s_A}$ of $1.9~\mathrm{Mpc}$ ($\approx 1.3\%$) and $2.2~\mathrm{Mpc}$  ($\approx 1.5\%$) for \zs~and \zp,  with $k_{max} = 0.1~h\rm{Mpc}^{-1}$ for the redshift bin centred at $z=0.9$}, in reasonable 
agreement with the error bars represented in Fig.~\ref{fig:psfit}. 
It should be stressed that the uncertainties given by the toy-program should be 
considered as a lower limit of the true error bars, as the galaxy power spectrum and the one without 
BAO oscillations, as well as the damping function $\eta(k)$  are supposed to be perfectly known. However,
the comparison between \zs~and \zp~remains meaningful. 

We have used this toy-program to compute $ \sigma_{s_A}  $ for an LSST-like survey 
and an effective survey area $\Omega_{surv}=10,000~\mathrm{deg^2}$, in six redshift bins in the range [0.4 - 2.2].
The results are summarised in Tab.~\ref{tab_err_sbao}, where one can see the increase of the error on the BAO scale 
due to \zp. Despite high galaxy number density, the relatively large value of
$ \sigma_{s_A}  $  around $z \sim 0.5$ can be explained by the more limited survey volume 
and larger \zp~dispersions, which has an even larger impact due to smaller Hubble parameter $H(z)$. 
The BAO scale would be rather well constrained by the LSST survey in the redshift range $ 0.7 \lesssim z \lesssim 1.6$, 
thanks to good photo-z reconstruction and large galaxy number densities. The degradation of BAO scale 
determination is visible for redshifts above $z \sim 1.6$ and even more clearly above $z \sim 2$, where   
$\sigma_{s_A} $ is increased by a factor {\col $\approx4$} in the case of \zp, compared 
to a full \zs.
 
We have also included forecasts of $ \sigma_{s_A}  $ for a fiducial spectroscopic survey with realistic galaxy number densities
in Tab.~\ref{tab_err_sbao}. Spectroscopic surveys naturally have a significantly lower galaxy number densities than photometric surveys
which limit their statistical power to map the LSS. {\col  To determine the galaxy number density for our fiducial spectroscopic survey, we have used the foreseen number of objects in the DESI survey~\citet{DESI2016}}. They expect to reach $\sim1625$ objects per 
$\mathrm {deg}^2$ up to redshift $\sim2.1$, by considering LRG's up to redshift 1 $(0.4 < z < 1.0)$, ELG's  
in the redshift range $0.6 < z < 1.6$ and QSO's up to redshift $z \sim2.1$. This galaxy density is about 
75 times less than the one expected for LSST ($\gtrsim 125000  \mathrm{~per~ deg^2}$). However, to be conservative, 
we have applied a factor 50 to the LSST galaxy number densities for the first 4 redshift bins, up to $z < 1.6$, 
and a factor 30 for the last two redshift bins $1.6 < z < 2.2$ to obtain the galaxy densities $\bar{n}_{gal}$ for 
computing forecasts for {\col our fiducial spectroscopic survey. }

\begin{table*}
\begin{tabular}{c||r|r|r||r|r|r||r|r|r|r}
redshift & \multicolumn{3}{c||}{ideal survey} & \multicolumn{3}{c||}{photometric survey} & \multicolumn{4}{c}{spectroscopic survey} \\ 
$ [z_{min},z_{max}]  $ & $\bar{n}_{gal}  $ &   \multicolumn{2}{c||}{$\sigma_{s_A}  [\%]$ } & $\bar{n}_{gal}  $ &   \multicolumn{2}{c||}{$\sigma_{s_A}  [\%]$ } &   $\bar{n}_{gal}$  & \multicolumn{3}{c} {$\sigma_{s_A}  [\%]$}  \\
                                &      & $k_{max1}$  & $k_{max2}$ &      & $k_{max1}$  & $k_{max2}$ &   & $k_{max1}$  & $k_{max2}$ & $k_{max3}$\\

\hline 
\hline 
$ [0.4-0.7]  $ & 1500        & 1.1    & 0.7 & 1500         & 1.2  & 0.7 &   30  & 1.3  & 0.9 & 0.5\\
$ [0.7-1.0]  $ & 1000    & 1.0   & 0.6 & 1000     & 1.2  & 0.7  &   20 & 1.4  & 0.9 & 0.6 \\
$ [1.0-1.3]  $ & 600     & 0.9    & 0.6 & 600       & 1.0 & 0.6  &   12 & 1.6  & 1.1 & 0.7 \\
$ [1.3-1.6]  $ & 300     & 0.9      & 0.5  & 300          & 1.1  & 0.8  &     6 & 2.9   & 2.0 &1.4 \\
$ [1.6-1.9]  $ & 130   & 0.9     & 0.6  & 130      & 1.6  & 1.3  &    4   & 4.9  & 4.0 & 2.8 \\
$ [1.9-2.2]  $ & 60     & 1.1   & 0.7  & 60      & 4.2  & 3.4 &    2  & 11.5  & 10.0 & 9.0\\
\hline 
\hline
\end{tabular}
\caption{{\col BAO scale determination relative uncertainty $\sigma_\mathrm{s_A}$ for different redshift ranges, 
for an ideal survey (left), for an LSST-like photometric survey (middle), and for a fiducial spectroscopic survey (right) with $\Omega_{surv}=10,000~\mathrm{deg^2}$. 
$\bar{n}_{gal}$ is the galaxy number density in units of $10^{-5} \, \mathrm{Mpc^{-3}}$. 
The \zs~redshifts are used for the ideal and the spectroscopic survey, and \zp~with errors similar to the reconstructed \zp~with the \bdtn~quality cut are used for the photometric survey. The columns giving $\sigma_{s_A}$ differ by the $k$-range used: $k_{max1}$ means $k \leq 0.07~\rm{Mpc^{-1}}$, $k_{max2}$ means $k \leq 0.1~\rm{Mpc^{-1}}$ and $k_{max3}$ means $k \leq 0.2~\rm{Mpc^{-1}}$ .}}
\label{tab_err_sbao}
\end{table*}

{\col The amount of information available in the high $k$-modes can be evaluated by comparing the gain in the BAO scale precision by extending the $k$-range in the fit, {\it i.e.} by fixing $k_{max}$ to 0.1 instead of 0.07~Mpc$^{-1}$. Such an extension would lead to a decrease of the relative uncertainty $\sigma_{s_A}$ by almost a factor of 2. It is probably slightly less if the natural smoothing produced by non-linear effects is taken into account. The gain is more limited at higher redshifts as the shot-noise becomes dominant. The ability to properly model the non-linear part of the LSS power spectrum may give access to a still larger $k$-range. A hint of the improvement is given by the last column, for a fiducial spectroscopic survey, with oscillations fitted up to $k = 0.2$~Mpc$^{-1}$. Note that such small scales are mainly whipped out, hidden by the shot-noise, if \zp~are used (see Fig.~\ref{fig:error}). 

Comparing the columns of Tab.~\ref{tab_err_sbao} shows the robustness of the BAO scale as a standard ruler, even if a relatively small fraction of the galaxies is used or if the redshift suffers from some dispersion. This parameter is statistically mostly immune to these defects below $z \approx 1.5$. Nevertheless, the LSS power spectrum is significantly damped if \zp~are used. So using the whole shape, and not only the BAO scale, would require a very precise knowledge of the \zp~properties at the origin of the damping function $\eta(k)$, while it is more directly provided by \zs~surveys. 

Here, only the isotropic BAO scale has been considered. However, it should be kept in mind that the BAO probe is more powerful in spectroscopic surveys, as they will be able to determine the BAO scale in the transverse and radial directions independently, and to measure the redshift space distortions, while LSST will be, mostly, sensitive to the transverse BAO scale due to \zp~smearing.  
}

%%%%%%%%%%%%%%%%%%%%%%%%%%%%%%%%%%%%%%%%%%%%%%%%%%%%%%%%%%%%%%%%%%%%%%%%%%%%%%%%%%%%%%
%%%%%%%%%%%%%%%%%%%%%%%%%%%%%%%%%%%%%%%%%%%%%%%%%%%%%%%%%%%%%%%%%%%%%%%%%%%%%%%%%%%%%%
\section{Conclusion and perspectives}
\label{sec:ccl}

The determination of accurate photometric redshifts of galaxies for redshifts up to 2, 
at least, is one the main challenges of the LSST survey. We have evaluated the 
impact of realistic \zp~uncertainties on the LSS power spectrum and the BAO scale uncertainty. 
The number density of usable galaxies and its evolution 
with redshift is one of the main parameters with major impact on cosmology with LSST.
We have determined the expected galaxy number density in the LSST survey, which decreases 
from $\sim 0.015\ \mathrm{Mpc^{-3}}$ at redshift $z \sim 1$ to $\sim 0.0015\ \mathrm{Mpc^{-3}}$  
at redshift $z \sim 1$ and associated \zp~errors. We have shown that LSST should be able 
to recover the BAO scale with few percents uncertainty up to redshift $z \lesssim 1.5$.

Around $z = 1$, \zp~are accurate, with low bias and there is a small fraction of
outliers. Therefore, LSST should be able to map accurately the galaxy distribution at this crucial 
stage of the history of the Universe, when the dark energy starts to dominate over matter.
The redshift of closer or more distant galaxies is less accurately measured in LSST, 
due to degeneracies between redshift and galaxy type.  
At low redshifts, the use of additional information like a BDT cut is helpful, leaving enough galaxies 
to properly compute the matter power spectrum and extract the BAO scale. At high redshifts
($z \gtrsim 1.5-2$), the observed galaxy number density is small and cannot compensate any more the \zp~smearing and the BAO signal is mainly washed out. 
{\col Our studies indicate that LSST will be able to recover with precision the BAO scale up to $z \approx 1.5$, with a \zp~reconstruction performances within requirements, confirming previous studies {\it e.g.}~\citet{Zhan2018}. However, our study also shows  that  
previous conclusions were rather optimistic regarding higher redshifts, as the recovered LSS power spectrum suffers larger 
uncertainties, due to the combination of a lower galaxy number density and poorer \zp~reconstruction. }

Using spherical shells and spherical harmonic transform instead of Euclidian grids and Fourier transform
is better suited to the analysis of galaxy surveys with large sky coverage. In addition, the theoretical 
framework for LSS analysis with spherical shells has been extensively developed in recent years, for instance \citet{Bonvin}, \citet{Lanusse}, \citet{angpow}.
We plan to carry a study of the impact of \zp~with 
separate determination of the radial and transverse BAO scales using spherical shells. 
At low redshift, specially around $z \sim 0.5$, the LSS power spectrum uncertainty is not limited by shot-noise 
in LSST, but suffers from \zp~spread and outliers. Identifying a population of galaxies with reliable 
\zp~at low redshifts is another useful follow-up study.

%%%%%%%%%%%%%%%%%%%%%%%%%%%%%%%%%%%%%%%%%%%%%%%%%%%%%%%%%%%%%%%%%%%%%%%%%%%%%%%%%%%%%%
%%%%%%%%%%%%%%%%%%%%%%%%%%%%%%%%%%%%%%%%%%%%%%%%%%%%%%%%%%%%%%%%%%%%%%%%%%%%%%%%%%%%%%
\begin{acknowledgements}
  Authors thanks the DESC Large-Scale Structures working group, especially D. Alonso and A. Slosar, for their useful comments. Authors thanks M. Moneuse for fruitful discussions and uses of CAMEL,  even if the cosmological parameters part exceeds the scope of this paper.
\end{acknowledgements}

\appendix

\section{Fit of the global shape of a power spectrum}

\begin{figure}[htpb]
 \includegraphics[width=\columnwidth]{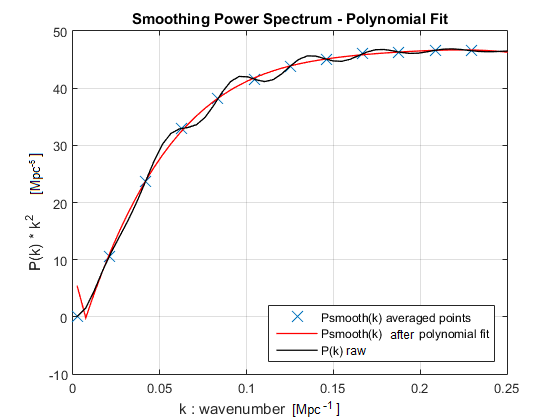}
\caption{Example of a power spectrum and its fitted baseline. The initial power spectrum is drawn in black, the inflexion points are indicated by the blue crosses and the red curve is the result of the fit, providing the global shape of the input power spectrum.}
\label{fig:baseline}
\end{figure}

{\col The oscillating spectrum $P_{{\rm wiggle}}$ is fitted by the usual ``Wiggle only''
method, which aims at recovering 
the BAO scale $s_A$.  

The polynomial description of the global shape $B$ is determined as follows:}
\begin{itemize} 
\item computing abscissas of the inflexion points where no contribution of baryonic oscillation is expected - they depend on the assumed value of
 the BAO scale,
\item deriving from data the ordinates of these points by averaging the power spectrum around these abscissas,
\item fitting this set of inflexion points by a polynomial of fifth degree.

\end{itemize} 
This method is illustrated by Fig.~\ref{fig:baseline}. We have used power spectra weighted by the square of the spatial wave number $(P(k) \times k^2)$ to enhance the readability. 
The baseline derived by this method is smooth by construction and adapt itself to include all kinds of damping. 

 This method has been validated with results obtained from grids filled with galaxies derived from matter
distribution unaffected by baryonic oscillations. Such a power spectrum, with
baryons but without baryonic oscillation, are available in the
Eisenstein approximation \citet{eisenstein}.
 
%%%%%%%%%%%%%%%%%%%%%%%%%%%%%%%%%%%%%%%%%%%%%%%%%%%%%%%%%%%%%%%%%%%%%%%%%%%%%%%%%%%%%%
%%%%%%%%%%%%%%%%%%%%%%%%%%%%%%%%%%%%%%%%%%%%%%%%%%%%%%%%%%%%%%%%%%%%%%%%%%%%%%%%%%%%%%
\bibliographystyle{aa}
\bibliography{biblio}
\end{document}